\documentclass[journal]{IEEEtran}
\usepackage[T1]{fontenc}
\usepackage[utf8]{inputenc}
\usepackage{xcolor}
\usepackage{bm}
\usepackage{amsmath}
\usepackage{amsthm}
\usepackage{amssymb}
\usepackage{graphicx}
\usepackage{setspace}
\PassOptionsToPackage{normalem}{ulem}
\usepackage{ulem}
\usepackage[unicode=true,
 bookmarks=true,bookmarksnumbered=true,bookmarksopen=true,bookmarksopenlevel=1,
 breaklinks=false,pdfborder={0 0 0},pdfborderstyle={},backref=false,colorlinks=false]
 {hyperref}
\hypersetup{pdftitle={Your Title},
 pdfauthor={Your Name},
 pdfpagelayout=OneColumn, pdfnewwindow=true, pdfstartview=XYZ, plainpages=false}

\makeatletter

\providecolor{lyxadded}{rgb}{0,0,1}
\providecolor{lyxdeleted}{rgb}{1,0,0}
\DeclareRobustCommand{\mklyxadded}[1]{\textcolor{lyxadded}\bgroup#1\egroup}
\DeclareRobustCommand{\mklyxdeleted}[1]{\textcolor{lyxdeleted}\bgroup\mklyxsout{#1}\egroup}
\DeclareRobustCommand{\mklyxsout}[1]{\ifx\\#1\else\sout{#1}\fi}

\theoremstyle{plain}
\newtheorem{thm}{\protect\theoremname}
\theoremstyle{remark}
\newtheorem{rem}[thm]{\protect\remarkname}

\usepackage[caption=false,font=footnotesize]{subfig}
\usepackage[utf8]{inputenc}
\usepackage{diagbox}
\usepackage{algorithm}
\usepackage{algorithmic}
\usepackage{color}
\usepackage{amsmath}
\usepackage{tikz}
\usetikzlibrary{calc}
\usepackage{bm}
\usepackage{graphicx}
\usepackage{textcomp}
\usepackage{booktabs}

\usepackage{amsthm}

\ifdefined\showcaptionsetup
 \PassOptionsToPackage{caption=false}{subfig}
\fi
\usepackage{subfig}
\makeatother

\providecommand{\remarkname}{Remark}
\providecommand{\theoremname}{Theorem}

\begin{document}
\title{Sensing for Free: Learn to Localize More Sources than Antennas without Pilots}
\author{Wentao~Yu,~\IEEEmembership{Member,~IEEE}, Khaled~B.~Letaief,~\IEEEmembership{Fellow,~IEEE},
and Lizhong~Zheng,~\IEEEmembership{Fellow,~IEEE}\thanks{This manuscript was accepted by the IEEE Journal on Selected Areas
in Communications (JSAC) on Jan. 5, 2026.}\thanks{This work was supported in part by the Hong Kong Research Grant Council
under Grant No. 16209023 and the Area of Excellence (AoE) Scheme Grant
No. AoE/E-601/22-R. \textit{(Corresponding author: Khaled B. Letaief)}}\thanks{Wentao Yu is with the Department of Electronic and Computer Engineering,
The Hong Kong University of Science and Technology, Kowloon, Hong
Kong, and also with the EECS Department, Massachusetts Institute of
Technology, Cambridge, MA 02139, USA (e-mail: wyuaq@connect.ust.hk).}\thanks{Khaled B. Letaief is with the Department of Electronic and Computer
Engineering, The Hong Kong University of Science and Technology, Kowloon,
Hong Kong (e-mail: eekhaled@ust.hk).}\thanks{Lizhong Zheng is with the EECS Department, Massachusetts Institute
of Technology, Cambridge, MA 02139, USA (e-mail: lizhong@mit.edu).}}
\maketitle
\begin{abstract}
Integrated sensing and communication (ISAC) represents a key paradigm
for future wireless networks. However, existing approaches often require
waveform modifications, dedicated pilots, or additional overhead that
complicates standards integration. We propose ``sensing for free''—performing
multi-source localization without pilots by reusing random and unknown
uplink data symbols, where sensing happens simultaneously with data
transmission, making it directly compatible with the 3GPP 5G NR and
6G specifications. With the ever-increasing number of devices in dense
6G networks, this approach becomes particularly compelling when combined
with sparse arrays, which can localize a much larger number of sources
compared to uniform arrays through the enlarged virtual array. However,
existing pilot-free multi-source localization algorithms for sparse
arrays have numerous drawbacks. They mostly first reconstruct an extended
covariance matrix and then apply subspace methods, which incur prohibitive
cubic complexity while being limited to second-order statistics. Performance
degrades under non-Gaussian modulated data symbols in cellular wireless
networks as the higher-order statistics that could further enhance
the localization capability remain unexploited. We address these challenges
with an attention-only transformer that directly processes raw signal
snapshots for grid-less end-to-end direction-of-arrival (DOA) estimation.
The model efficiently captures higher-order statistics while being
permutation-invariant and adaptive to varying numbers of snapshots.
Our algorithm greatly outperforms state-of-the-art artificial intelligence
(AI)-based benchmarks with over $30\times$ reduction in parameters
and runtime, and enjoys excellent generalization under practical mismatches.
In addition, it can effectively handle multipath propagation and mixed
modulation types. Beyond localization, our algorithm can also enhance
multi-user MIMO beam training through angular reciprocity. The estimated
DOAs in the uplink data transmission stage can significantly prune
downlink beam sweeping candidates and enhance system throughput via
sensing-assisted beam management. Overall, this work demonstrates
how reusing existing random data payloads for sensing can enhance
both multi-source localization and beam management, two key AI-for-communication
use cases in 3GPP efforts towards 6G.
\end{abstract}


\section{Introduction}

\subsection{Background}

Integrated sensing and communication (ISAC) has emerged as a key paradigm
for 6G wireless networks, attracting significant interest across academia,
industry, and also standardization bodies \cite{liu2022integrated,xie2025sensing,itu_r_m2160_2023,5GAmericas2024,3gpp_tr_22_837,chen20235g,lin2025tale}.
For instance, ISAC is recognized as one of the six major use cases
of 6G \cite{itu_r_m2160_2023}, and has been actively studied by the
3rd generation partnership project (3GPP), e.g., TR 22.837 in Release 19
and the ongoing discussions for Release 20 \cite{3gpp_tr_22_837,chen20235g,lin2025tale}.
The basic idea is to enable the communication infrastructure itself
to perform sensing functions, typically by repurposing the communication
signals for radar-like tasks. Existing ISAC approaches, however, often
require modifications to the waveforms or frame structures to embed
sensing capabilities, which can undermine the communication performance
or demand upgrades to the existing infrastructure. By contrast, we
are interested in a scenario where sensing can be achieved ``for free''
during regular data transmissions, without using additional pilots
or radio resources. In particular, we focus on the uplink of a cellular
system, where a multi-antenna base station (BS) attempts to localize
multiple sources by processing the unknown uplink data symbols. This
type of ISAC would allow next-generation networks to gain situational
awareness with no extra overhead, making it easy for practical deployment.

\begin{figure}[t]
\centering{}\includegraphics[width=0.45\textwidth]{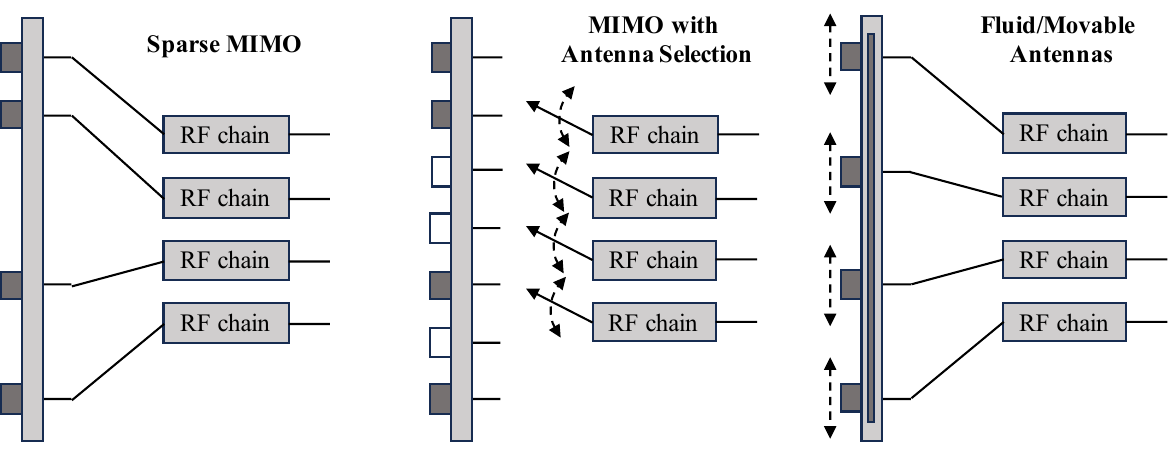}\caption{SLAs and their variants, including but not limited to MIMO systems
with antenna selection \cite{Molisch2004MIMO} and fluid/movable antennas
\cite{Wong2021Fluid,Zhu2024Modeling}. \label{fig:Generalized_sparse_MIMO}}
\end{figure}

At the same time, the demand to localize a large number of sources
has become increasingly prominent, driven by the proliferation of
massive internet of things (IoT) \cite{nguyen2022internet}, unmanned
aerial vehicles (UAV) communications \cite{zeng2016wireless}, and
low-altitude economy \cite{jiang2025integrated}. Using a conventional
$M$-element uniform linear array (ULA), subspace methods such as
multiple signal classification (MUSIC) can only resolve at most $M-1$
uncorrelated sources since the noise subspace must be non-trivial
\cite{schmidt1986multiple}. To overcome such a limitation, sparse
linear arrays (SLAs) have been proposed, such as nested \cite{pal2010nested},
co-prime \cite{vaidyanathan2011sparse}, and minimum redundancy arrays
(MRAs) \cite{moffet1986minimum}, which sparsely deploy antennas over
a large aperture to enhance the sensing capability. A properly designed
$M$-antenna SLA can form a much larger virtual array, enabling the
localization of $O(M^{2})$ uncorrelated sources using extended covariance
matrix, far more than the number of antennas $M$ \cite{li2025sparse-linear}.
These advantages have motivated interest in considering SLAs for ISAC
systems \cite{li2025sparse}. In \cite{min2025integrated}, the authors
found that by exploiting the virtual array for localization and physical
array for communications, SLAs can achieve both enhanced sensing capability
and competitive communication performance. Some recent investigations
also focused on the advantages of SLAs in enhancing communication
performance \cite{wang2023can,zhou2024sparse,chen2024near}. In fact,
SLAs represent both established and emerging paradigms in wireless
communications. In MIMO systems with antenna selection \cite{Molisch2004MIMO}
or switch-based hybrid analog-digital beamforming \cite{zhang2020hybrid},
activating a subset of antennas effectively creates an SLA. Moreover,
recently emerging flexible-antenna technologies, such as fluid \cite{Wong2021Fluid}
and movable \cite{Zhu2024Modeling} antenna systems, deploy antennas
in flexible positions within a large aperture, and can also be naturally
configured as SLAs to exploit their advantages in sensing and localization.
The algorithms studied in this work are relevant to different practical
scenarios\footnote{For fluid/movable antennas, antenna positions are usually set at the
start of each coherence interval and then kept fixed during data transmission
\cite{wong2023slow,liao2025joint}. The algorithms operate on a block
of $T$ snapshots within that interval, which is similar to static
SLAs.}. We show the variants of SLAs in Fig. \ref{fig:Generalized_sparse_MIMO}.

In this paper, we study multi-source localization with SLAs using
unknown uplink data symbols and ask the following question: \textit{Can
a base station localize more sources than its number of antennas during
regular data transmission, without pilots or extra resources?} We
investigate DL-based algorithms that provide affirmative answers to
this question that are accurate, fast, and robust. We further showcase
the fundamental value of such ``sensing for free” capability for wireless
system design, in particular beam management.

\subsection{Related Works}

\subsubsection{Multi-Source Localization with SLAs}

While pilot-free ISAC with sparse arrays is an emerging idea in wireless
communications \cite{li2025sparse}, a closely related problem, i.e.,
sparse array DOA estimation, has been extensively studied in array
signal processing. Classical subspace-based multi-source localization
algorithms, including MUSIC \cite{schmidt1986multiple} and Root-MUSIC
\cite{barabell1983improving}, rely on eigen-decomposition of the
sample spatial covariance matrix (SCM) to separate signal and noise
subspaces. For SLAs, the key challenge is reconstructing the missing
entries of the SCM to exploit the larger virtual aperture, which can
be formulated as a constrained optimization problem using the maximum
likelihood principle. Optimization techniques, such as semi-definite
programming (SDP) \cite{yang2014discretization}, nuclear norm \cite{li2015off}
and atomic norm \cite{tang2014near} minimization, Wasserstein distance
minimization \cite{wang2019grid}, etc., have been proposed. Recently,
DL methods have also been studied to reconstruct the extended SCM
for subsequent subspace methods. Most of them use convolutional neural
networks (CNNs) as the backbone. In \cite{barthelme2021DoA}, leveraging
the semidefinite property of the SCM, the authors proposed using CNNs
to output auxiliary matrices whose Gramians approximate the extended
SCM. In \cite{wu2022gridless}, the authors proposed estimating the
first row to construct Toeplitz-structured SCMs, exploiting the inherent
structure of array response matrices. In \cite{chen2025subspace},
a subspace representation learning approach was proposed that prioritizes
signal subspace preservation over exact SCM reconstruction. These
DL-based methods employ various loss functions including Frobenius
norm \cite{barthelme2021DoA,wu2022gridless}, affine invariant distance
\cite{barthelme2021DoA} and subspace distance \cite{chen2025subspace},
such as principal angles between subspaces \cite{bjorck1973numerical}.

While these constitute important related works, they represent signal
processing approaches that differ fundamentally from the problem addressed
in this study. First, these methods are predicated on Gaussian signal
assumptions and rely exclusively on second-order statistics, specifically
the SCM. Although the SCM provides sufficient statistics for Gaussian
processes, modulated data symbols in practical communication systems
exhibit non-Gaussian characteristics, leading to suboptimal performance
when applying these approaches. Second, computational complexity presents
a significant limitation. These methods employ a two-stage pipeline
comprising SCM reconstruction followed by classical subspace methods,
necessitating eigen-decomposition of the SCM and exhaustive spectrum
search to separate signal and noise subspaces. This results in prohibitive
computational complexity scaling as $O(N^{3})$, where $N\gg M$ represents
the number of elements the virtual array corresponding to the SLA.
Alternative approaches that attempt to exploit higher-order statistics,
such as MUSIC-like algorithms leveraging fourth-order cumulants, incur
an even more prohibitive complexity of $O(N^{6})$ \cite{yuen1997DOA,peng2024under}.
Third, prior works typically assume the availability of hundreds of
signal snapshots for DOA estimation, which is unrealistic given the
short coherence times in communication systems. Under practical constraints
of limited snapshots, the sample SCM becomes unreliable and as a result
leads to performance degradation of subspace methods.

Parallel to developments from the algorithm side, recent non-asymptotic
analysis also provides finite-sample guarantees related to DOA estimation
with sparse arrays. In \cite{sarangi2023super}, Sarangi \textit{et
al.} established a precise trade-off between temporal and spatial
measurements using high-dimensional concentration bounds, and importantly,
dispelled the common belief that super-resolution with virtual arrays
intrinsically demands a very large number of snapshots, which was
followed by \cite{shahsavari2025cramer} that presented a non-asymptotic
Cramér-Rao bound (CRB) analysis under a total-power constraint across
antennas. These results show that accurate sparse array DOA estimation
can be achieved with a practical number of snapshots available in
cellular wireless networks.

\subsubsection{Sensing with Random Communication Signals}

Frame structures of modern wireless communication systems devote only
a small fraction of time-frequency resources to deterministic and
known pilots, while most of the frame consists of random and unknown
data symbols. Some recent works studied the theoretical benefits of
combining those random signals with deterministic pilots for different
sensing problems. In \cite{xie2025bistatic}, the authors formulated
bistatic target detection with both deterministic pilots and random
payloads and derived a generalized likelihood ratio test (GLRT)-based
detector with an asymptotic false-alarm rate analysis. In \cite{xu2025exploiting},
the authors studied monostatic ISAC using both pilots and random data
for sensing, and studied an ISAC precoder optimization problem. In
\cite{graff2024ofdm}, the authors derived a Ziv-Zakai bound for orthogonal
frequency division multiplexing (OFDM) time-of-arrival estimation
using both pilots and unknown data, and quantified the performance
gains over pilot-only sensing. These theoretical studies have shown
the potentials of random communication signals for enhancing pilot-based
sensing.

\subsubsection{Sensing-Assisted Beam Management}

Angular reciprocity in massive MIMO systems enables sensing-assisted
beam management by leveraging the shared angular sparsity between
uplink and downlink channels, which is observed even in frequency
division duplex (FDD) systems \cite{hugl2002spatial,imtiaz2014directional}.
In \cite{zhang2018directional}, the authors demonstrated directional
training frameworks leveraging this reciprocity through measurement
results. While they did not study specific algorithms for uplink DOA
estimation, the results laid the foundation of sensing-assisted beam
training. Recent advances in multimodal ISAC have further explored
richer sensing paradigms, integrating vision, radio, and radar data
for reducing the overhead and improving the accuracy of beam prediction
\cite{alrabeiah2020millimeter,charan2022vision,zhang2025multimodal}.

While these works have established the fundamentals and benefits of
sensing-assisted beam management, they either rely on dedicated pilots
for angular parameter estimation or require additional sensing modalities
and corresponding infrastructure upgrades. This leaves an opportunity
for ``sensing for free'' approaches that can achieve sensing-assisted
beam management without requiring additional resources or system modifications.

\subsection{Contributions}

The contributions of this paper are mainly two-fold, which are summarized
from system and algorithmic perspectives.

From a system standpoint, we demonstrate that, by leveraging unknown
uplink data symbols already present in each frame, multi-source localization
can be achieved at the same time with data transmission, without any
dedicated overhead. Because the method leaves frame structures and
waveforms untouched, it could be easily embedded into forthcoming
wireless systems. Exploiting the enlarged virtual aperture of SLAs,
the BS can accurately localize the DOAs for more users than physical
antennas. These DOA estimates, in turn, are accurate enough to enable
a significant pruning of the downlink beam sweeping candidates, remarkably
reducing beam management overhead and enhancing system throughput.
The sensing-assisted beam management is also shown to work robustly
under uplink-downlink reciprocity mismatch. The results in our study
motivate a rethinking of how uplink data symbols can be better harnessed
to facilitate the sensing functionality.

From an algorithmic perspective, conventional multi-source localization
algorithms for SLAs rely on covariance reconstruction followed by
subspace methods. They are limited to second-order statistics, computationally
expensive and prone to performance loss under the non-Gaussian data
symbols and the limited snapshots. We circumvent subspace methods
with an attention-only transformer that processes raw signal snapshots
of SLAs to efficiently capture higher-order statistics for grid-less
end-to-end DOA estimation, avoiding the high-complexity covariance
reconstruction and decomposition steps. Our algorithm consistently
outperforms state-of-the-art benchmarks across a wide signal-to-noise-ratio
(SNR) range while offering a significant reduction of over $30\times$
in computational complexity under typical system settings. It also
generalizes under various practical mismatches, scales gracefully
to large-scale MIMO arrays, and effectively handles multipath propagation
and mixed modulation types. Extensive simulation results confirm these
advantages, establishing a novel paradigm for pilot-free multi-source
localization with sparse arrays. Ablation studies are also provided
to investigate the reasons behind the performance gains.

To avoid repetition later, we summarize the scope of this work in
the following remark.
\begin{rem}[``sensing for free''\label{rem:sense-free}]
\textit{ Throughout this paper, the BS opportunistically reuses the
random and unknown uplink data payloads to perform multi-source localization
in parallel with data transmission, without introducing any new pilots,
signaling, or frame changes. The resulting uplink DOA estimates serve
as side information to prune the downlink beam sweeping candidates,
thereby reducing beam management overhead and improving system throughput.}
\end{rem}

\subsection{Paper Organization and Notation}

The remaining parts of this article are organized as follows. In Section
\ref{sec:Preliminaries}, we discuss the preliminaries of the system
model and define the problem. In Section \ref{sec:Prior-Art}, we
introduce the prior arts of multi-source localization with sparse
arrays and their limitations. In Section \ref{sec:Proposed-Snap-TF},
we propose the Snap-TF algorithm and discuss its properties and insights.
In Section \ref{sec:Simulation-Results}, extensive simulation results
are presented to demonstrate the advantages of the proposed algorithm
from various aspects. In Section \ref{sec:Further-Application:-Sensing-Ass},
we discuss the further application of the proposed algorithm to sensing-assisted
beam management of multi-user MIMO (MU-MIMO) systems. Lastly, we conclude
the paper in Section \ref{sec:Conclusion-and-Future}.

\textit{Notation:} Throughout this work, we denote scalars by italic
lowercase letters, column vectors with boldface lowercase letters,
while matrices by boldface uppercase letters, respectively. For any
vector $\mathbf{a}$, $[\mathbf{a}]_{1:i}$ denotes its first $i$
elements. For any matrix $\mathbf{A}$, $\mathbf{A}^{*}$, $\mathbf{A}^{\mathsf{T}}$,
$\mathbf{A}^{\mathsf{H}}$, $[\mathbf{A}]_{:,1:i}$, and $[\mathbf{A}]_{i,j}$
are its conjugate, transpose, conjugate transpose, the first $i$
columns, and the $(i,j)$-th element, respectively. The Euclidean
and Frobenius norms are written as $\lVert\mathbf{a}\rVert_{2}$ and
$\lVert\mathbf{A}\rVert_{F}$, respectively, and $|\cdot|$ is the
cardinality of a set. The operators $\operatorname{diag}(\cdot)$,
$\Re({\cdot})$, and $\Im({\cdot})$ generate a diagonal matrix and
extract the real and imaginary parts, respectively. The imaginary
unit is $j$. The natural logarithm is denoted by $\ln(\cdot)$.

\section{Preliminaries \label{sec:Preliminaries}}

\subsection{System Model with Sparse MIMO Arrays}

\begin{figure}[t]
\centering
\includegraphics[width=0.322\textwidth]{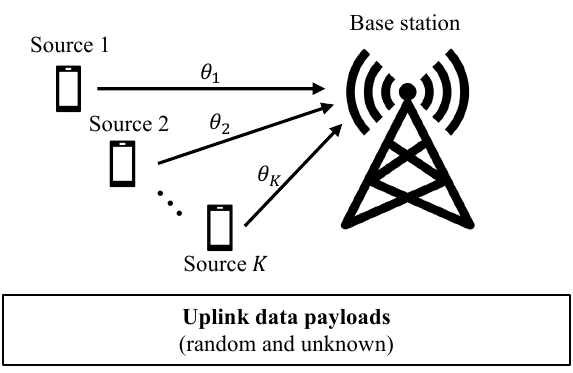}\caption{System model. In the uplink, there are $K$ sources impinging upon
a BS from DOAs $\bm{\theta}=\{\theta_{1},\theta_{2},\ldots,\theta_{K}\}$.
The BS aims to localize these sources based on the received random
and unknown uplink data payloads. \label{fig:System-model.}}
\end{figure}

Assuming that there are $K$ narrowband and far-field sources $\{s_{k}\}_{k=1}^{K}$
with a carrier wavelength $\lambda$ impinging upon a BS with a linear
array from DOAs $\bm{\theta}=\{\theta_{1},\theta_{2},\ldots,\theta_{K}\}$
in the uplink data transmission stage, as shown in Fig. \ref{fig:System-model.}.
We assume the DOAs are independent and identically distributed (i.i.d.)
and are drawn randomly from a uniform distribution $\mathcal{U}(\theta_{\min},\theta_{\max})$,
where $0\leq\theta_{\min}\leq\theta_{\max}\leq\pi$ are the minimum
and maximum values of the angle range, respectively. The DOAs also
satisfy a minimum angle separation constraint defined as $\Delta_{\bm{\theta},\text{min}}\triangleq\min_{i\ne j}\left|\theta_{i}-\theta_{j}\right|,\forall i,j\in\{1,2,\ldots,K\}$.
The array is equipped with $M$ antennas with an array aperture of
$Nd$ where $M\leq N$. It can either be uniform or sparse\footnote{While we focus on SLAs, the proposed algorithms also apply to uniform
arrays. This work focuses on algorithms; see \cite{li2025sparse-linear}
for sparse array design.}, where the antenna positions are chosen from the integer multiples
of half the wavelength $d=\frac{\lambda}{2}$. We use $\bm{\Omega}=\{\Omega_{1},\Omega_{2},\ldots,\Omega_{M}\}$
to denote the indices of antennas, where $\Omega_{m}$ is an integer
representing that the position of the $m$-th antenna is $\Omega_{m}d$.
In general, we configure that $1=\Omega_{1}<\Omega_{2}<\ldots<\Omega_{M}=N$.
For the special case of ULAs, we have $M=N$ and $\bm{\Omega}=\{1,2,\ldots,N\}$.
Fig. \ref{fig:three_arrays_aligned} shows the system model and a
comparison of uniform and sparse arrays.

The signals are transmitted from the sources to the array for $T$
snapshots. For each snapshot index $t\in\{1,2,\ldots,T\}$, the received
signal at the array, i.e., $\mathbf{y}_{\bm{\Omega}}(t)\in\mathbb{C}^{M\times1}$,
is
\begin{equation}
\begin{aligned}\mathbf{y}_{\bm{\Omega}}(t) & =\sum_{k=1}^{K}p_{k}\mathbf{a}_{\bm{\Omega}}(\theta_{k})s_{k}(t)+\mathbf{n}_{\bm{\Omega}}(t)\\
 & =\mathbf{A}_{\bm{\Omega}}(\bm{\theta})\mathbf{P}\mathbf{\mathbf{s}}(t)+\mathbf{n}_{\bm{\Omega}}(t)\\
 & =\mathbf{H}_{\bm{\Omega}}\mathbf{\mathbf{s}}(t)+\mathbf{n}_{\bm{\Omega}}(t),
\end{aligned}
\label{eq:system-model}
\end{equation}
where $p_{k}\in\mathbb{C}$, $s_{k}(t)\in\mathbb{C}$, and $\mathbf{a}_{\bm{\Omega}}(\theta_{k})\in\mathbb{C}^{M\times1}$
are the path loss, the source signal, and the array response vector
of the $k$-th source, respectively. In addition, $\mathbf{A}_{\bm{\Omega}}(\bm{\theta})=[\mathbf{a}_{\bm{\Omega}}(\theta_{1}),\mathbf{a}_{\bm{\Omega}}(\theta_{2}),\ldots,\mathbf{a}_{\bm{\Omega}}(\theta_{K})]\text{\ensuremath{\in\mathbb{C}^{M\times K}}}$
is the array response matrix, $\mathbf{P}=\operatorname{diag}(p_{1},p_{2},\ldots,p_{K})\in\mathbb{C}^{K\times K}$
is the matrix form of the path loss, and $\mathbf{s}(t)=[s_{1}(t),s_{2}(t),\ldots,s_{K}(t)]^{\mathsf{T}}\in\mathbb{C}^{K\times1}$
is the collection of source signals. The multi-user MIMO channel of
the SLA is denoted by $\mathbf{H}_{\bm{\Omega}}=\mathbf{A}_{\bm{\Omega}}(\bm{\theta})\mathbf{P}\in\mathbb{C}^{M\times K}$.
The noise is given by $\mathbf{n}_{\bm{\Omega}}(t)\in\mathbb{C}^{M\times1}$.
The source signals $\mathbf{s}(t)$ and the noise $\mathbf{n}_{\bm{\Omega}}(t)$
are independent of each other, and they are both temporally uncorrelated
across different snapshots. The array response vector $\mathbf{\mathbf{a}}_{\bm{\Omega}}(\theta_{k})$
is 
\begin{equation}
\mathbf{\mathbf{a}}_{\bm{\Omega}}(\theta_{k})=\left[e^{j\pi(\Omega_{1}-1)\sin(\theta_{k})},\ldots,e^{j\pi(\Omega_{M}-1)\sin(\theta_{k})}\right]^{\mathsf{T}}.
\end{equation}
The elements of the SLA array response vector $\mathbf{\mathbf{a}}_{\bm{\Omega}}(\theta_{k})$
are selected from those of the ULA array response vector $\mathbf{\mathbf{a}}(\theta_{k})$
with $\mathbf{\mathbf{a}}_{\bm{\Omega}}(\theta_{k})=\bm{\Gamma}_{\bm{\Omega}}\mathbf{\mathbf{a}}(\theta_{k})$,
where $\mathbf{\mathbf{a}}(\theta_{k})=[1,e^{j\pi\sin(\theta_{k})},\ldots,e^{j\pi(N-1)\sin(\theta_{k})}]$
and $\bm{\Gamma}_{\bm{\Omega}}\in\mathbb{R}^{M\times N}$ is a row
selection matrix whose elements are ones at the $\Omega_{m}$-th position
of the $m$-th row and zeros elsewhere. Similarly, we have $\mathbf{A}_{\bm{\Omega}}(\bm{\theta})=\bm{\Gamma}_{\bm{\Omega}}\mathbf{A}(\bm{\theta})$,
with $\mathbf{A}(\bm{\theta})=[\mathbf{a}(\theta_{1}),\mathbf{a}(\theta_{2}),\ldots,\mathbf{a}(\theta_{K})]\text{\ensuremath{\in\mathbb{C}^{N\times K}}}$
being the ULA array response matrix, and $\mathbf{y}_{\bm{\Omega}}(t)=\bm{\Gamma}_{\bm{\Omega}}\mathbf{y}(t)$,
where $\mathbf{y}(t)\in\mathbb{C}^{N\times1}$ denotes the received
signals by the ULA. We make the following assumptions on the source
signals and the noise:
\begin{itemize}
\item \textbf{Source signals} ${s_{k}(t)}$ are all zero-mean random processes.
We consider both Gaussian and non-Gaussian signals, and assume that
each $s_{k}(t)$ is independent identically distributed (i.i.d.) across
snapshots. The signal $s_{k}(t)$ can either be circular complex Gaussian
or practical non-Gaussian modulated symbols, such as QPSK (quadrature
phase shift keying) and 16QAM (quadrature modulation), etc. We assume
that the source signals are mutually independent\footnote{The proposed algorithm in this work is applicable to not only independent
sources but also coherent sources arising from multipath propagation
\cite{yuen1997DOA}. The case of coherent sources will be discussed
separately in Section \ref{subsec:Coherence-Sources-Arising}.} and also uncorrelated with the noise, if not otherwise stated. \textit{We
consider the case where the source signals }$s_{k}(t)$ \textit{are
unknown during the estimation process, corresponding to unknown uplink
data symbols. }Hence, our multi-source DOA estimation happens simultaneously
with uplink data transmission, and requires no additional overhead
and no modifications to the frame structures or waveforms.
\item \textbf{Noise} $\mathbf{n}_{\bm{\Omega}}(t)$ is assumed zero-mean
Gaussian distributed and independent across snapshots, following $\mathcal{CN}(\mathbf{0},\eta\mathbf{I})$,
where $\eta$ is the noise power. 
\begin{figure}[t]
\begin{singlespace}
\subfloat{\includegraphics{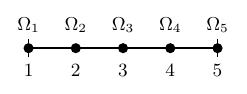}}\\
\subfloat{\includegraphics{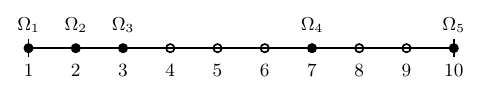}}\caption{An comparison of 5-element uniform and sparse MIMO arrays. The top
array is a ULA with $M=N=5$, and the bottom one is a special SLA,
called minimum redundancy array (MRA) \cite{moffet1986minimum}, with
$M=5$ and $N=10$. The black circles denote the positions where the
antennas are placed. \label{fig:three_arrays_aligned}}
\end{singlespace}
\end{figure}
\end{itemize}
After $T$ snapshots\footnote{We assume DOAs are unchanged within a short block of at least $T$
symbols. In the simulations, we use $T=50$ by default and also report
results with value as small as $T=10$. The proposed method does not
rely on long continuous traffic and remains applicable when only brief
bursts exist. For very sparse or sporadic traffic, activity detection
and temporal aggregation could be combined as an extension, which
we leave for future work.}, we stack the received signal snapshots into the matrix form. For
the full $N$-element ULA, we have $\mathbf{Y}=[\mathbf{y}_{\bm{}}(1),\mathbf{y}_{\bm{}}(2),\ldots,\mathbf{y}_{\bm{}}(T)]\in\mathbb{C}^{N\times T}$,
while for the $M$-element SLA, we obtain $\mathbf{Y}_{\bm{\Omega}}=[\mathbf{y}_{\bm{\Omega}}(1),\mathbf{y}_{\bm{\Omega}}(2),\ldots,\mathbf{y}_{\bm{\Omega}}(T)]\in\mathbb{C}^{M\times T},$
where $\mathbf{Y}_{\bm{\Omega}}=\bm{\Gamma}_{\bm{\Omega}}\mathbf{Y}$.
We next define the noiseless spatial covariance matrices (SCMs). The
full $N$-element ULA has noiseless SCM $\mathbf{R}=\mathbf{A}(\bm{\theta})\mathbf{P}\mathbf{A}^{\mathsf{H}}(\bm{\theta})\in\mathbb{C}^{N\times N}$.
The corresponding noiseless SCM for the $M$-element SLA is $\mathbf{R}_{\bm{\Omega}}=\bm{\Gamma}_{\bm{\Omega}}\mathbf{R}\bm{\Gamma}_{\bm{\Omega}}^{\mathsf{T}}\in\mathbb{C}^{M\times M}$.
In practice, the sample SCMs of SLA and ULA are obtained based on
the noisy received signal snapshots, which are respectively given
by $\tilde{\mathbf{R}}_{\bm{\Omega}}\triangleq\frac{1}{T}\mathbf{Y}_{\bm{\Omega}}\mathbf{Y}_{\bm{\Omega}}^{\mathsf{H}}$
and $\tilde{\mathbf{R}}\triangleq\frac{1}{T}\mathbf{Y}\mathbf{Y}^{\mathsf{H}}$.

\subsection{Problem Statement}

Based on $T$ snapshots of received signals at the $M$-element SLA,
i.e., $\mathbf{Y}_{\bm{\Omega}}$, the target is to estimate the DOAs
of $K$ sources, i.e., $\bm{\theta}$, in which $K$ could be larger
than $M$. We work under the ``sensing for free'' setup (see Remark
\ref{rem:sense-free}). As the true and estimated DOAs are two sets,
to remove the permutation ambiguity when comparing their distance,
we define the mean square error (MSE) of the estimated angles $\hat{\bm{\theta}}$
and the ground-truth $\bm{\theta}$ as
\begin{equation}
\operatorname{MSE}(\hat{\boldsymbol{\theta}},\boldsymbol{\theta})=\frac{1}{K}\min_{\boldsymbol{\Pi}_{K}\in\mathcal{P}_{K}}\left\Vert \boldsymbol{\Pi}_{K}\hat{\boldsymbol{\theta}}-\boldsymbol{\theta}\right\Vert _{2}^{2},\label{eq:MSE}
\end{equation}
where $\boldsymbol{\Pi}_{K}$ represents a specific $K\times K$ permutation
matrix while $\mathcal{P}_{K}$ denotes the set of all possible $K\times K$
permutation matrices. The minimum can be easily obtained by sorting
both $\hat{\boldsymbol{\theta}}$ and ${\boldsymbol{\theta}}$ according
to the rearrangement inequality \cite{hardy1934inequalities}, which
is widely used to evaluate the DOA estimation performance.

\section{Prior Art \label{sec:Prior-Art}}

While there are no previous works for multi-source localization with
sparse arrays using unknown data symbols, we summarize the state-of-the-art
DL-based algorithms for sparse array DOA estimation using Gaussian
signals, which will serve as important benchmarks later. The dataset
is defined as $\{\mathbf{Y}_{\bm{\Omega}}^{(i)},K^{(i)},\bm{\theta}^{(i)}\}_{i=1}^{I}$,
where $I$ is the number of samples in the dataset and the superscript
$(\cdot)^{(i)}$ denotes the $i$-th sample. For the $i$-th data
sample, the DOA estimation algorithm takes both the received signal
snapshots at the SLA, i.e., $\mathbf{Y}_{\bm{\Omega}}^{(i)}\in\mathbb{C}^{M\times T}$,
and the number of sources, i.e., $K^{(i)}$ as inputs, while the outputs
are the estimated DOAs whose ground-truth labels are $\bm{\theta}^{(i)}\in\mathbb{R}^{K\times1}$.

Most previous methods follow a two-stage pipeline. First, they learn
to reconstruct the noiseless SCM of the corresponding virtual ULA,
i.e., ${\mathbf{R}}^{(i)}$, from the sample SCM of the ULA, i.e.,
$\tilde{\mathbf{R}}_{\bm{\Omega}}^{(i)}$, both of which can be obtained
based on the dataset as described earlier. Second, they apply subspace
methods, e.g., MUSIC or Root-MUSIC, for DOA estimation \cite{barthelme2021DoA,wu2022gridless,chen2025subspace}.
These approaches employ convolutional neural networks (CNNs) as backbones,
denoted by $f_{\mathcal{W}}(\cdot)$ with $\mathcal{W}$ being the
network parameters, and assume that the transmitted signals $\mathbf{s}(t)$
are Gaussian distributed. The main differences of existing DL-based
subspace methods lies in the model output and loss function, as listed
in Table \ref{tab:method_comp}.

In the following, we begin by summarizing the two key steps of the
existing algorithms, i.e., covariance reconstruction and subspace
methods, before discussing their limitations for multi-source localization
with unknown uplink symbols. 
\begin{table}[t]
\begin{centering}
\caption{Existing learning methods for sparse array DOA estimation \label{tab:method_comp}}
\par\end{centering}
\centering{}\footnotesize \begin{tabular}{@{\hskip 1pt}l@{\hskip 4pt}l@{\hskip 4pt}l@{\hskip 1pt}}     \toprule     \textbf{Method} & \textbf{Model output} & \textbf{Loss function} \\     \midrule     DCR-G-Fro \cite{barthelme2021DoA} &       $\hat{\mathbf{R}}        =f_{\mathcal{W}}(\tilde{\mathbf{R}}_{\bm{\Omega}})         f_{\mathcal{W}}^{\mathsf{H}}(\tilde{\mathbf{R}}_{\bm{\Omega}})$ &       $\|\hat{\mathbf{R}}-\mathbf{R}\|_{F}$ \\[2pt]     DCR-G-Aff \cite{barthelme2021DoA} &       $\hat{\mathbf{R}}        =f_{\mathcal{W}}(\tilde{\mathbf{R}}_{\bm{\Omega}})         f_{\mathcal{W}}^{\mathsf{H}}(\tilde{\mathbf{R}}_{\bm{\Omega}})$ &       $\bigl\|         \ln\bigl(           \mathbf{R}^{-\tfrac12}\,\hat{\mathbf{R}}\,\mathbf{R}^{-\tfrac12}         \bigr)       \bigr\|_{F}$ \\[2pt]     DCR-T \cite{wu2022gridless} &       $\hat{\mathbf{v}}        =f_{\mathcal{W}}(\tilde{\mathbf{R}}_{\bm{\Omega}}),         \hat{\mathbf{R}}        =\operatorname{Toep}(\hat{\mathbf{v}})$ &       $\bigl\|         \hat{\mathbf{v}}          -[\mathbf{A}(\bm{\theta})\mathbf{A}^{\mathsf{H}}(\bm{\theta})]_{1,:}^{\mathsf{T}}       \bigr\|_{2}$ \\[2pt]     SRL \cite{chen2025subspace} &       $\hat{\mathbf{R}}        =f_{\mathcal{W}}(\tilde{\mathbf{R}}_{\bm{\Omega}})         f_{\mathcal{W}}^{\mathsf{H}}(\tilde{\mathbf{R}}_{\bm{\Omega}})$ &       $\mathrm{dist}_{\mathrm{sub}}\bigl(\hat{\mathbf{R}},\mathbf{R}\bigr)$ \\     \bottomrule   \end{tabular}
\end{table}

\subsection{Covariance Reconstruction}

Building upon the intuition that the noiseless SCM of the ULA should
be both Hermitian and positive semi-definite, the authors of \cite{barthelme2021DoA}
trained a CNN to estimate an auxiliary matrix $f_{\mathcal{W}}(\tilde{\mathbf{R}}_{\bm{\Omega}})\in\mathbb{C}^{N\times N}$,
whose Gramian $\hat{\mathbf{R}}=f_{\mathcal{W}}(\tilde{\mathbf{R}}_{\bm{\Omega}})f_{\mathcal{W}}^{\mathsf{H}}(\tilde{\mathbf{R}}_{\bm{\Omega}})$
is an estimate of $\mathbf{R}$, which will be used in subspace methods.
To measure the performance of deep covariance reconstruction (DCR),
the authors have proposed two loss functions, i.e., the Frobenius
norm and the affine invariant distance, called DCR-G-Fro and DCR-G-Aff,
respectively.

Noticing that the SCM of ULA is a Toeplitz matrix characterized by
its first row, the authors of \cite{wu2022gridless} proposed to estimate
it using a CNN $f_{\mathcal{W}}(\cdot)$ whose output is $\hat{\mathbf{\mathbf{v}}}=f_{\mathcal{W}}(\tilde{\mathbf{R}}_{\bm{\Omega}})\in\mathbb{C}^{N\times1}$.
The loss function is chosen as the Euclidean norm between $\hat{\mathbf{\mathbf{v}}}$
and $[\mathbf{A}(\bm{\theta})\mathbf{A}^{\mathsf{H}}(\bm{\theta})]_{1,:}^{\mathsf{T}}$.
The estimated SCM is then constructed as a Toeplitz matrix according
to the estimated first row, i.e., $\hat{\mathbf{R}}=\operatorname{Toep}(\hat{\mathbf{v}})$,
which will be used in subspace methods. This algorithm is shortened
as DCR-T.

The authors of \cite{chen2025subspace} observed that subspace methods
require only that the reconstructed SCM $\hat{\mathbf{R}}=f_{\mathcal{W}}(\tilde{\mathbf{R}}_{\bm{\Omega}})f_{\mathcal{W}}^{\mathsf{H}}(\tilde{\mathbf{R}}_{\bm{\Omega}})$
can share the same signal subspace as the noiseless ULA SCM $\mathbf{R}$.
Exact recovery of $\mathbf{R}$ is a harder problem that is not necessary.
Based on this, they defined various metrics that instead measure the
distances between the signal subspaces of $\hat{\mathbf{R}}$ and
$\mathbf{R}$, denoted by $\operatorname{dist_{sub}}(\hat{\mathbf{R}},\mathbf{R})$.
We denote the eigen-decomposition of $\hat{\mathbf{R}}$ as 
\begin{equation}
\hat{\mathbf{R}}=\begin{bmatrix}\hat{\mathbf{U}} & \hat{\mathbf{V}}\end{bmatrix}\begin{bmatrix}\hat{\bm{\Lambda}}_{K}\\
 & \hat{\bm{\Lambda}}_{N-K}
\end{bmatrix}\begin{bmatrix}\hat{\mathbf{U}}^{\mathsf{H}}\\[0.3em]
\hat{\mathbf{V}}^{\mathsf{H}}
\end{bmatrix},\label{eq:EVD}
\end{equation}
where $\hat{\bm{\Lambda}}_{K}$ and $\hat{\bm{\Lambda}}_{N-K}$ are
both diagonal matrices consisting of the $K$ largest and the $(N-K)$
smallest eigen-values. The columns of $\hat{\mathbf{U}}\in\mathbb{C}^{N\times K}$
and $\hat{\mathbf{V}}\in\mathbb{C}^{N\times(N-K)}$ consist of the
eigen-vectors that correspond to the signal and the noise subspaces,
respectively. A similar decomposition can be performed for the noiseless
ULA SCM $\mathbf{R}$, which will yield $\mathbf{U}$ and $\mathbf{V}$.
The signal subspaces for the ground-truth and the estimated ULA SCMs
are the range spaces of $\mathbf{U}$ and $\hat{\mathbf{U}}$, respectively.
The authors used principle angles as a measure of the distance between
signal subspaces \cite{bjorck1973numerical}, given by
\begin{equation}
\operatorname{dist_{sub}(\hat{\mathbf{R}},\mathbf{R})}=\bigl\lVert\cos^{-1}\bigl(\sigma\bigl(\mathbf{U}^{\mathsf{H}}\hat{\mathbf{U}}\bigr)\bigr)\bigr\rVert_{2},
\end{equation}
where $\sigma(\cdot)$ returns the singular value vector of the input
matrix in descending order, and $\cos^{-1}(\cdot)$ is the element-wise
inverse cosine function. This algorithm is shortened as SRL, standing
for subspace representation learning.

\subsection{Subspace Methods}

The reconstructed SCM of the ULA, i.e., $\hat{\mathbf{R}}$, is then
fed into subspace methods, such as MUSIC \cite{schmidt1986multiple}
and root-MUSIC \cite{barabell1983improving}, to perform DOA estimation.
The initial step for subspace methods is an eigen-decomposition of
the reconstructed SCM $\hat{\mathbf{R}}\in\mathbb{C}^{N\times N}$
as shown in (\ref{eq:EVD}). This yields the signal subspace $\hat{\mathbf{U}}\in\mathbb{C}^{N\times K}$
and the noise subspace $\hat{\mathbf{V}}\in\mathbb{C}^{N\times(N-K)}$,
which should be orthogonal. Because the signal subspace is spanned
by the columns of the array response matrix $\mathbf{A}(\bm{\theta})$,
the corresponding array response vectors are therefore orthogonal
to the noise subspace, which implies that 
\begin{equation}
\mathbf{a}^{\mathsf{H}}(\theta_{k})\hat{\mathbf{V}}\hat{\mathbf{V}}^{\mathsf{H}}\mathbf{a}(\theta_{k})=\lVert\hat{\mathbf{V}}^{\mathsf{H}}\mathbf{a}(\theta_{k})\rVert_{2}^{2}=0,\label{eq:Basis}
\end{equation}
for $k\in\{1,2,\ldots,K\}$. This is the basis of subspace methods.
MUSIC tackles (\ref{eq:Basis}) via spectrum peak search \cite{schmidt1986multiple},
while Root-MUSIC instead resorts to solving a polynomial equation
\cite{barabell1983improving}.

\subsection{Limitations of Prior Art}

Existing DL-enhanced subspace methods, such as MUSIC \cite{schmidt1986multiple}
and Root-MUSIC \cite{barabell1983improving}, face significant computational
challenges in sparse MIMO systems. First, reconstructing the $N\times N$
virtual ULA SCM from the $M\times M$ SLA SCM requires a large-scale
neural network model. In addition, after covariance reconstruction,
these algorithms require eigen-decomposition of the $N\times N$ reconstructed
covariance matrix $\hat{\mathbf{R}}$ followed by exhaustive search
over candidate angles, with each step incurring $O(N^{3})$ complexity.
Since the extended array aperture $N$ can substantially exceed the
number of physical antennas $M$ in sparse arrays, this $O(N^{3})$
complexity becomes prohibitive for practical implementation.

To address the non-Gaussian nature of communication signals (e.g.,
QPSK or 16QAM), which exhibit non-zero higher-order cumulants, several
approaches have extended MUSIC-type algorithms to exploit higher-order
statistics \cite{yuen1997DOA,peng2024under}. However, these methods
construct and decompose an $N^{2}\times N^{2}$ fourth-order cumulant
matrix, escalating the computational complexity to an even more prohibitive
$O(N^{6})$. Furthermore, both second-order and higher-order approaches
suffer from performance degradation when operating with limited snapshots,
as the sample statistics become unreliable.

Motivated by these limitations, this work proposes a grid-less end-to-end
DOA estimation framework that directly maps raw signal snapshots $\mathbf{Y}_{\bm{\Omega}}\in\mathbb{C}^{M\times T}$
to estimated DOAs. This approach completely bypasses explicit covariance
reconstruction and subspace decomposition steps and maintains computational
efficiency for large-scale sparse MIMO arrays. Additionally, the proposed
neural architecture is designed to efficiently learn higher-order
signal statistics and adapt to varying numbers of snapshots, enhancing
the generalization performance of existing DL-enhanced methods.

\section{Our Proposed Snap-TF Algorithm \label{sec:Proposed-Snap-TF}}

In view of the limitations of existing approaches discussed above,
this work proposes pilot-free multi-source DOA estimation algorithms
designed for SLA systems based on modulated data symbols. The primary
objective is to develop an end-to-end DL-based solution that can effectively
capture higher-order statistics while circumventing the computational
burden of covariance reconstruction and decomposition as well as exhaustive
spectrum search inherent in existing DL-augmented subspace methods.

In the following, we first discuss the key requirements of neural
architecture to realize these goals, and then propose our transformer
on snapshots (Snap-TF) algorithm carefully designed to satisfy these
requirements.

\begin{algorithm*}[!t] \caption{Proposed Snap-TF: Permutation Invariant Transformer on Snapshots} \label{alg:multi_layer_transformer} \begin{algorithmic}[1] \STATE \textbf{Input:} Received signal snapshots $\mathbf{Y}_{\bm{\Omega}}\in\!\mathbb{R}^{M\times T}$, number of sources $K$, number of transformer layers $L$, width of attention activation $d_{\text{attn}}$ \STATE \textbf{Output:} Estimated DOAs $\hat{\bm{\theta}} \in \mathbb{R}^{K\times 1}$ \STATE \textbf{Trainable Parameters:} Weights $\mathbf{W}_1,\mathbf{W}_2,\mathbf{W}_3,\mathbf{W}_4,\mathbf{W}_Q^{(\ell)},\mathbf{W}_K^{(\ell)},\mathbf{W}_V^{(\ell)}$, and biases $\mathbf{b}_1$, $\mathbf{b}_2$, $\mathbf{b}_3$, $\mathbf{b}_4$ \STATE \textbf{Initialize:}  Initial feature set $\mathbf{S}^{(0)}\gets[\Re(\mathbf{Y}_{\bm{\Omega}}^{\mathsf{T}}),\Im(\mathbf{Y}_{\bm{\Omega}}^{\mathsf{T}})]^{\mathsf{T}}\in\!\mathbb{R}^{2M\times T}$ \FOR{$\ell=1$ \TO $L$}     \STATE \textbf{Self‑Attention Block:}     \vspace{0.3\baselineskip}     \STATE \quad $\displaystyle       \mathbf{Z}^{(\ell)}       \gets \mathrm{LayerNorm}\mkern1mu\Bigl(           \mathbf{S}^{(\ell-1)}           + {\smash{\overbrace{(\mathbf{W}_V^{(\ell)}\mathbf{S}^{(\ell-1)})}^{\scriptstyle\mathbf{V}^{(\ell)}\in\, \mathbb{R}^{d_{\text{attn}}\times T}}}\cdot \mathrm{Softmax}\mkern1mu\Bigl(               \frac{                 \smash{\overbrace{(\mathbf{W}_K^{(\ell)}\mathbf{S}^{(\ell-1)})^{\mathsf{T}}}^{\scriptstyle(\mathbf{K}^{(\ell)})^{\mathsf{T}}\in\, \mathbb{R}^{T \times d_{\text{attn}}}}} \cdot                 \smash{\overbrace{(\mathbf{W}_Q^{(\ell)}\mathbf{S}^{(\ell-1)})}^{\scriptstyle\mathbf{Q}^{(\ell)}\in\, \mathbb{R}^{d_{\text{attn}}\times T}}}               }{\sqrt{d_{\text{attn}}}}           }       \Bigr)$     \STATE \textbf{Feed‑Forward Block:}     \vspace{0.3\baselineskip}     \STATE \quad $\mathbf{S}^{(\ell)}\gets            \mathrm{LayerNorm} \mkern1mu\bigl(\mathbf{Z}^{(\ell)}+\mathbf{W}_2^{(\ell)}\,\mathrm{ReLU}\mkern1mu\bigl(\mathbf{W}_1^{(\ell)}\mathbf{Z}^{(\ell)}+\mathbf{b}_1^{(\ell)}\mathbf{1}^{\mathsf{T}}\bigr)            +\mathbf{b}_2^{(\ell)}\mathbf{1}^{\mathsf{T}}\bigr)$      \ENDFOR \vspace{0.3\baselineskip} \STATE \textbf{Global Mean Pooling:} \vspace{0.3\baselineskip} \STATE \quad $\mathbf{q}\gets \tfrac1T\sum_{t=1}^{T}\mathbf{s}_t^{(L)} \quad (\text{where} \,\, \mathbf{S}^{(L)}\triangleq[\mathbf{s}_1^{(L)},\mathbf{s}_2^{(L)},\ldots,\mathbf{s}_T^{(L)}])$ \vspace{0.3\baselineskip} \STATE \textbf{Output Block:} \vspace{0.3\baselineskip} \STATE \quad $\hat{\bm{\theta}}\gets            \bigl[\mathbf{W}_4\,\mathrm{ReLU}\mkern1mu\bigl(\mathbf{W}_3\mathbf{q}+\mathbf{b}_3\bigr)            +\mathbf{b}_4\bigr]_{1:K}$ \STATE \textbf{Return:} $\hat{\bm{\theta}}$ \end{algorithmic} \end{algorithm*}

\subsection{Key Requirements on Neural Architecture}

In general, the proposed Snap-TF must generalize across both the different
numbers of snapshots $T$ and the different orders of snapshots, while
being able to capture the higher-order statistics for accurate DOA
estimation. We introduce the key requirements that should be taken
into account as follows.

\subsubsection{Permutation Invariance}

The algorithm should be permutation invariant to the order of the
snapshots. This is because the received signals $\mathbf{y}_{\bm{\Omega}}(t)\in\mathbb{C}^{M\times1}$
are drawn i.i.d. across different snapshot $t$, as the DOAs are static
during estimation and the symbols and noise vectors are both i.i.d.
across different snapshots. This means that the order of the snapshots
should not affect the estimated DOAs. Denote the proposed Snap-TF
DOA estimation network as $g_{\mathbb{\mathcal{W}}}(\mathbf{\mathbf{Y}_{\bm{\Omega}}})$,
where $\mathcal{W}$ is the set of trainable parameters. By permutation
invariance, we refer to the property that permuting the columns of
$\mathbf{Y}_{\bm{\Omega}}$ (i.e., a permutation of the snapshot index)
will not change the output DOA estimates of the neural network, i.e.,
$g_{\mathbb{\mathcal{W}}}(\mathbf{Y}_{\bm{\Omega}})=g_{\mathbb{\mathcal{W}}}(\mathbf{Y}_{\bm{\Omega}}\bm{\Pi}_{T}),\forall\bm{\Pi}_{T}\in\mathcal{P}_{T}$,
where $\mathcal{P}_{T}$ denotes the set of all $T\times T$ permutation
matrices. Otherwise, the network will learn misleading information
regarding the order of the snapshots, and have poor generalization
performance.

Existing DL-augmented subspace methods introduced in the last section
inherently satisfy this requirement since their input is the sample
SCM $\tilde{\mathbf{R}}_{\bm{\Omega}}\in\mathbb{C}^{M\times M}$ whose
dimension remains constant regardless of the number of snapshots $T$,
and whose value is invariant with respect to a permutation of the
snapshot index $t$ in $\mathbf{y}_{\bm{\Omega}}(t)$. However, designing
end-to-end neural networks that can process raw signal snapshots to
capture higher-order statistics while maintaining similar generalization
capabilities is much more challenging, and requires careful architectural
considerations. Popular feed-forward and recurrent neural networks
(RNNs) are sensitive to the input order and hence not suitable for
the considered problem.

\subsubsection{Generalization to Varying Numbers of Snapshots}

In real-world deployments the number of snapshots $T$ is rarely fixed.
It shrinks when coherence times are short or processing budgets are
tight, and it grows when the coherence times are longer. A robust
DOA estimator should therefore generalize across arbitrary $T$, returning
accurate estimates whether it sees fewer or more snapshots than in
training. The natural way to meet this requirement is to treat the
received signal snapshots $\mathbf{Y}_{\bm{\Omega}}\in\mathbb{C}^{M\times T}$
as an unordered set, with each column $\mathbf{y}_{\bm{\Omega}}(t)\in\mathbb{C}^{M\times1}$
being an element of the set whose order is exchangeable with other
set elements. Consequently, the network must be permutation invariant
and able to process the input of different numbers of snapshots without
retraining.

\subsubsection{Higher-Order Interactions}

The higher-order statistics is helpful in expanding the number of
sources that an array can localize. For the second-order statistics,
consider an $M=5$ element MRA with $\bm{\Omega}=\{1,2,5,8,10\}$
and array response vector $\mathbf{a}_{\bm{\Omega}}(\theta)$. The
phase term reads $e^{j\pi(\Omega_{i}-1)\sin(\theta)}$, the entries
of the sample SCM $\tilde{\mathbf{R}}_{\bm{\Omega}}$ depends on $e^{j\pi(\Omega_{i}-\Omega_{j})\sin(\theta)},\forall i,j$,
or the difference $|\Omega_{i}-\Omega_{j}|\in\{0,1,2,\ldots,9\}$
with 10 elements. Hence, the difference co-array acts like a virtual
ULA with $N=10$ antennas, more than the number of physical antennas
$M=5$. By contrast, for the fourth-order statistics, the array response
vector $\mathbf{a}_{\bm{\Omega}}(\theta)$ is replaced by the Kronecker
product $\mathbf{a}_{\bm{\Omega}}(\theta)\otimes\mathbf{a}_{\bm{\Omega}}(\theta)$.
The phase term now reads $e^{j\pi(\Omega_{i}+\Omega_{j}-2)\sin(\theta)}$,
so every unordered pair $(i,j)$ behaves like a virtual antenna at
$(\Omega_{i}+\Omega_{j}-2),\forall i,j$ \cite{chevalier2005virtual}.
The resulting virtual array contains 14 distinct elements. This example
illustrates how higher-order statistics expand array localization
capability. Since modulated data symbols often exhibit non-zero fourth-order
statistics, leveraging this property enables localization of even
more sources than conventional covariance-based methods.

Despite these benefits, utilizing higher-order statistics in existing
subspace methods incurs prohibitive computational complexity as discussed
previously. The neural architecture we propose should therefore efficiently
capture higher-order statistics while avoiding these computational
bottlenecks. The received signal snapshots $\mathbf{Y}_{\bm{\Omega}}\in\mathbb{C}^{M\times T}$
can be treated as an unordered set of $\mathbf{y}_{\bm{\Omega}}(t)\in\mathbb{C}^{M\times1}$,
where temporal ordering is irrelevant. The higher-order statistics
are determined by higher-order interactions among these set elements.
Our neural architecture should effectively capture these interactions
at low computational cost, specifically avoiding the matrix decomposition
costs of $O(N^{3})$ of covariance-based subspace methods and $O(N^{6})$
of fourth-order approaches.

\subsection{Architecture of the Proposed Snap-TF}

\begin{figure*}[t]
\centering
\includegraphics[width=0.75\textwidth]{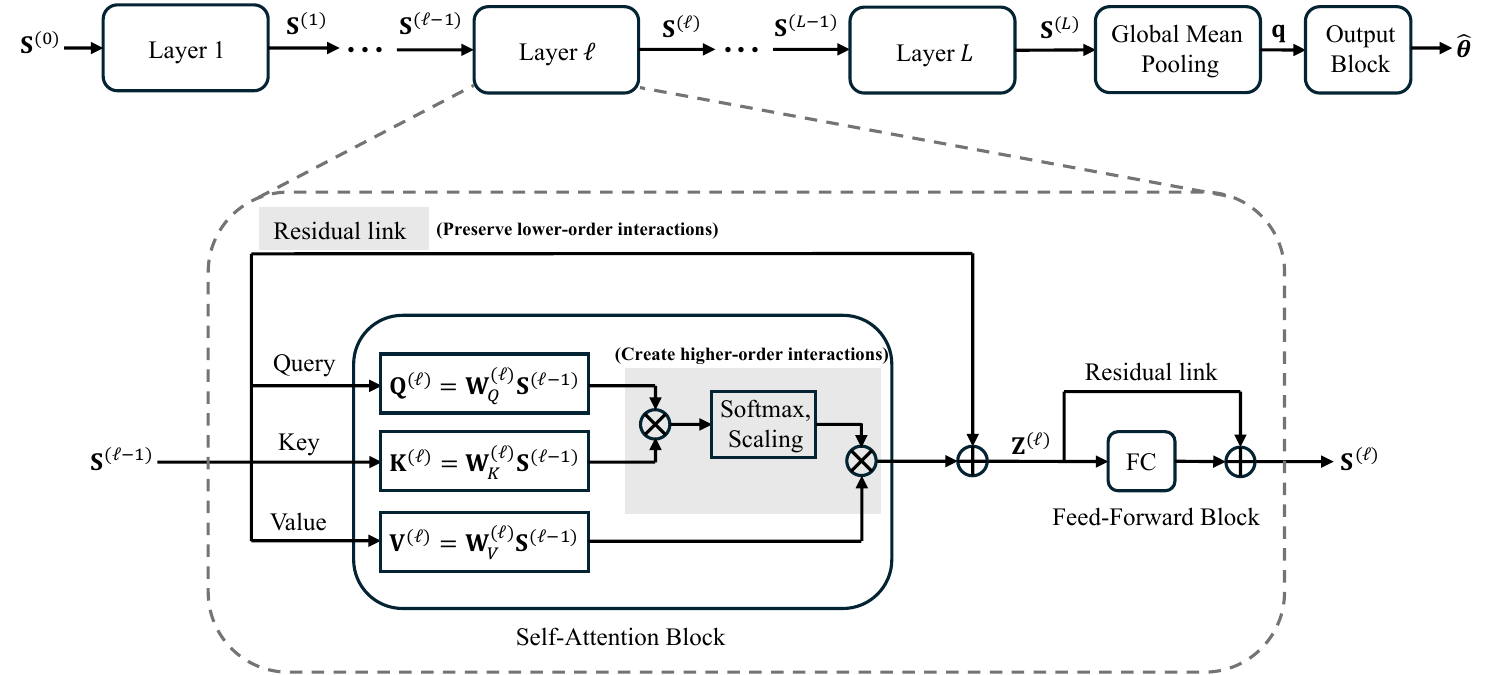}\caption{Schematic diagram of the proposed Snap-TF algorithm. \label{fig:Schematic-diagram-of}}
\end{figure*}

Next, we introduce the proposed Snap-TF architecture for multi-source
DOA estimation. Its pseudo-codes are shown in \textbf{Algorithm \ref{alg:multi_layer_transformer}},
while its schematic diagram is illustrated in Fig. \ref{fig:Schematic-diagram-of}.
We discuss the details as follows. 

\subsubsection{Signals as Tokens}

We initialize by treating each raw snapshot as an input token to the
transformer network. Specifically, for an $M$-element SLA, each snapshot
$\mathbf{y}_{\bm{\Omega}}(t)\in\mathbb{C}^{M\times1}$ is split into
its real and imaginary parts and concatenated into a real-valued $2M$-dimensional
vector. Stacking $T$ such vectors forms the initial feature set $\mathbf{S}^{(0)}\in\mathbb{R}^{2M\times T}$,
whose $t$-th column corresponds to the $t$-th snapshot. We bypass
the tokenization procedures in vanilla transformers \cite{vaswani2017attention}
to let the network directly process higher-order interactions between
raw signal snapshots to learn the higher-order statistics that are
important for DOA estimation under non-Gaussian modulated symbols.
Also, we deliberately drop the positional encoding and masking commonly
employed in vanilla transformers \cite{vaswani2017attention} to preserve
the permutation invariance of the architecture with respect to the
ordering of signal snapshots $\mathbf{y_{\bm{\Omega}}}(t)$. Since
a permutation of the snapshot index should not affect the output,
we treat $\mathbf{S}^{(0)}$ as the initial feature set whose columns
are the unordered set elements.

\subsubsection{Self-Attention Blocks for Higher-Order Interactions}

Modulated data symbols carry higher-order statistics that can enhance
localization capability, yet classical subspace methods require constructing
and decomposing large and potentially redundant cumulant matrices.
Moreover, the arrangement of higher-order terms directly determines
the achievable localization capability \cite{chevalier2006high}.
In practice, with a limited number of snapshots, the optimal arrangement
depends on many factors, calling for adaptive and data-driven computation.
Higher-order cumulants are permutation-invariant symmetric polynomials
of the set elements of $\mathbf{S}^{(0)}$. Since self-attention can
effectively approximate such polynomials with theoretical guarantees
\cite{takeshita2025approximation}, stacked self-attention layers
in Snap-TF can implicitly and adaptively approximate the necessary
higher-order information without explicitly reconstructing cumulant
matrices, yielding compact features for the downstream DOA estimator.

Snap-TF employs a stack of single-head self-attention blocks to capture
the higher-order interactions among the raw signal snapshots. We denote
the feature set after the $\ell$-th layer as $\mathbf{S}^{(\ell)}\triangleq[\mathbf{s}_{1}^{(\ell)},\mathbf{s}_{2}^{(\ell)},\ldots,\mathbf{s}_{T}^{(\ell)}]$,
which is also an unordered set. The self-attention block, $\mathbb{R}^{2M\times T}\to\mathbb{R}^{2M\times T}$,
takes the feature set $\mathbf{S}^{(\ell)}\in\mathbb{R}^{2M\times T}$
as input and is defined in line 7 of \textbf{Algorithm \ref{alg:multi_layer_transformer}},
where $\mathbf{W}_{Q}^{(\ell)},\mathbf{W}_{K}^{(\ell)},\mathbf{W}_{V}^{(\ell)}$
are the trainable parameter matrices for $\ell=1,2,\ldots,L$ with
$d_{\text{attn}}$ being the width of attention activation. In the
$\ell$-th layer, the query, key, and value are respectively denoted
by $\mathbf{Q}^{(\ell)},\mathbf{K}^{(\ell)},\mathbf{V}^{(\ell)}\in\mathbb{R}^{d_{\text{attn}}\times T}$.
In the self-attention block, each element of the feature set, i.e.,
$\mathbf{s}_{t}^{(\ell)},\forall t\in\{1,2,\ldots T\}$, attends to
all other elements in $\mathbf{S}^{(\ell)}$, generating attention
weights $\mathrm{Softmax}(\frac{(\mathbf{K}^{(\ell)})^{\mathsf{T}}\mathbf{Q}^{(\ell)}}{\sqrt{d_{\text{attn}}}})$
that encode the pairwise statistical relationships of the set elements
of $\mathbf{S}^{(\ell)}$. For example, define $\mathbf{q}_{i}^{(\ell)}=\mathbf{W}_{Q}^{(\ell)}\mathbf{s}_{i}^{(\ell)}$
and $\mathbf{k}_{j}^{(\ell)}=\mathbf{W}_{K}^{(\ell)}\mathbf{s}_{j}^{(\ell)}$.
Their attention weight is $\mathrm{softmax}((\mathbf{k}_{j}^{(\ell)})^{\mathsf{T}}\mathbf{q}_{i}^{(\ell)}/\sqrt{d_{\mathrm{attn}}})$
which depends on the two set elements $\mathbf{s}_{i}^{(\ell)}$ and
$\mathbf{s}_{j}^{(\ell)}$, and is thus adaptive to the input of the
model. Self-attention hence creates adaptively weighted higher-order
interactions without explicitly constructing cumulant matrices. In
addition, we also adopt layer normalization \cite{ba2016layer}, i.e.,
$\mathrm{LayerNorm}(\cdot)$, and residual connection \cite{he2016deep}.
It is also followed by a feed-forward block in line 9, which are fully-connected
(FC) layers to to enhance the representation capability, with $\mathrm{ReLU(\cdot)=\max(0,\cdot)}$
being the rectified linear unit activation function.

Through the stacking of $L$ self-attention layers, Snap-TF progressively
constructs higher-order interactions. The first layer output encodes
pairwise interactions, which, when processed by subsequent layers,
enables the formation of terms involving up to four elements, and
so forth. In addition, as seen from Fig. \ref{fig:Schematic-diagram-of},
while the self-attention operation creates higher-order interactions,
the residual link preserves the lower-order interactions computed
in previous layer. As a result, $L$ stacked attention layers can
capture interactions of order up to $2^{L}$ and preserves all lower-order
information, so $L=\lceil\log_{2}(k+1)\rceil$ layers are enough to
capture $k$-th order interactions. For example, $L=3$ layers can
provide sufficient representational capacity for DOA estimation under
non-Gaussian modulated symbols used in communications, whose non-Gaussianity
is well characterized by cumulants up to the eighth order \cite{chevalier2005virtual}.
This will be verified by the ablation studies in Section \ref{subsec:Ablation-Studies},
suggesting that very few layers of Snap-TF can already offer competitive
performance.

Another important property of the self-attention and feed-forward
blocks in line 5-10 is the inherent permutation equivariance. Permutations
of the columns in feature set $\mathbf{S}^{(\ell)}$ produce correspondingly
permuted outputs with unchanged individual values. This property arises
because self-attention relies exclusively on inner products between
columns of the query $\mathbf{Q}^{(\ell)}$ and the key $\mathbf{K}^{(\ell)}$
matrices, disregarding temporal ordering and treating the $T$ snapshots
as an unordered set. Consequently, applying global pooling to the
feature set elements yields a permutation-invariant mapping from raw
signal snapshots to estimated DOAs, as will be introduced below.

\subsubsection{Global Mean Pooling and Output Block}

After $L$ layers, we obtain the final feature set of $T$ elements
$\mathbf{S}^{(L)}=[\mathbf{s}_{1}^{(L)},\mathbf{s}_{2}^{(L)},\ldots,\mathbf{s}_{T}^{(L)}]$.
To produce DOA estimates, we aggregate these $T$ features into a
fixed-size vector using global mean pooling, i.e., computing the mean
across all columns of $\mathbf{S}^{(L)}$. This pooled feature vector
$\mathbf{q}$ is then fed into a small FC network for final DOA estimation.
Global mean pooling ensures that the Snap-TF DOA estimator $g_{\mathcal{W}}(\cdot)$
satisfies permutation invariance. The mean pooling is column-symmetric,
so the order of the columns of $\mathbf{S}^{(L)}$ does not change
the result of pooling. Together with the permutation equivariance
property of self-attention blocks, overall, Snap-TF's output is invariant
to permutations of the snapshots, formally guaranteeing $g_{\mathbb{\mathcal{W}}}(\mathbf{Y}_{\bm{\Omega}})=g_{\mathbb{\mathcal{W}}}(\mathbf{Y}_{\bm{\Omega}}\bm{\Pi}_{T}),\forall\bm{\Pi}_{T}\in\mathcal{P}_{T}$,
where $\mathcal{P}_{T}$ here denotes the set of all $T\times T$
permutation matrices. Global mean pooling also enables generalization
to different numbers of snapshots $T$. The network can be trained
on one range of $T$ values and applied to scenarios with different
$T$ without retraining, since pooling can naturally aggregate different
numbers of snapshots.

After global mean pooling, we apply FC layers to the feature vector
$\mathbf{q}$ to estimate the DOAs. To handle the varying number of
sources\footnote{In 3GPP NR uplink, the base station either schedules uplink data on
the physical uplink shared channel (PUSCH) by sending downlink control
information (DCI) on the physical downlink control channel (PDCCH)
addressed to a specific user via a radio network temporary identifier
(RNTI), or enables configured-grant resources by radio resource control
\cite{TS38212,TS38214,TS38213,TS38211}. Hence, the number of sources
$K$ is available at the BS. \label{fn:In-3GPP-NR}} $K$ in practical systems, we set the output dimension to the maximum
allowed number of sources $K_{\max}$. The model outputs $K_{\max}$
angle estimates, and we select the first $K$ elements corresponding
to the actual number of sources $K$, which enables the model to accommodate
varying numbers of sources without requiring architectural changes.

\subsection{Computational Complexity}

If $k$-th order statistics are considered, conventional subspace
methods generally need to construct an $N^{\frac{k}{2}}\times N^{\frac{k}{2}}$
cumulant matrix to exploit the extended virtual array, e.g., $N\times N$
for covariance ($k=2$) and $N^{2}\times N^{2}$ for fourth-order
cumulants ($k=4$), assuming $k$ is an even number \cite{chevalier2005virtual}.
The MUSIC-like algorithms first need to reconstruct such a cumulant
matrix based on the received signal snapshots of the SLA, which often
requires a very large-scale CNN. Then, they perform eigen-decomposition
of the $N^{\frac{k}{2}}\times N^{\frac{k}{2}}$ cumulant matrix and
apply spectrum search to separate the signal and noise subspaces,
which incur a prohibitive complexity of $O(N^{\frac{3k}{2}})$ where
$N\gg M$. The computational complexity is prohibitive especially
for large arrays and non-Gaussian modulated symbols that call for
higher-order statistics. For example, MRAs with $M=28$ physical antennas
have $N=244$ virtual elements \cite{schwartau2021large}.

Unlike subspace methods, the proposed Snap-TF algorithm completely
eliminates not only the computationally expensive covariance (or cumulant
matrix) reconstruction step, but also the prohibitive eigen-decomposition
and exhaustive spectrum search steps. It only needs to process the
raw received signal snapshots of SLAs with dimension $M$. This complexity
reduction is particularly significant for large-scale SLAs. By contrast,
computational bottleneck of the proposed Snap-TF algorithm lies in
computing attention matrices $\mathbf{K}^{(\ell)\mathsf{T}}\mathbf{Q}^{(\ell)}$
with complexity $O(T^{2}d_{\text{attn}})$, where $d_{\text{attn}}$
is a hyper-parameter often set slightly larger than $N$. The complexity
of Snap-TF algorithm is roughly linear with respect to $N$. Also,
the number of snapshots $T$ in practical communication systems is
usually small due to the short coherence interval. In addition, techniques
to accelerate the attention step is extensively studied in the literature,
e.g., linear attention \cite{wang2020linformer}, which scales linearly
with $T$ and can further reduce the complexity. More importantly,
Snap-TF enables the algorithm to capture the $k$-th order statistics
using $L=\text{\ensuremath{\lceil\log_{2}(k+1)\rceil}}$ layers. Hence,
the overall complexity is roughly $O(\lceil\log_{2}(k+1)\rceil T^{2}d_{\text{attn}})$,
typically smaller than existing DL-based subspace methods introduced
in Section \ref{sec:Prior-Art}. The advantage is particularly large
for large-scale arrays, small number of snapshots $T$, and when higher-order
statistics are used.

In Section \ref{subsec:Complexity}, we will compare the complexity
in terms of both the trainable parameters and the runtime between
the proposed and benchmark methods to show Snap-TF's substantial computational
advantages in both aspects.

\section{Simulation Results\label{sec:Simulation-Results}}

In this section, we conduct simulations to compare the proposed Snap-TF
algorithm with the state-of-the-art DL-based benchmarks introduced
in Section \ref{sec:Prior-Art}. We will first introduce simulation
settings and then discuss the results.

\subsection{Settings \label{subsec:Settings}}

\subsubsection{System Settings\label{subsec:System-Settings}}

We consider the ``sensing for free'' setup as introduced in Remark
\ref{rem:sense-free}. Specifically, we first consider a multi-user
MIMO system where the BS employs an SLA with $M=5$ physical antennas
and index set $\bm{\Omega}=\{1,2,5,8,10\}$, which is a 5-element
MRA\footnote{We use MRAs to illustrate simulation results without loss of generality.
The studied algorithms are geometry-agnostic as they exploit the subspace
structure of received signal statistics, which holds for any array
geometry. Different geometries yield different effective degree of
freedom through their unique lags, but the algorithmic principles
remain broadly applicable.} \cite{moffet1986minimum} with an aperture of $N=10$ half-wavelength
units. Its difference co-array will lead to a 10-element virtual ULA
that can localize $K_{\max}=N-1=9$ sources using subspace methods.
In addition, to illustrate the scalability of the proposed Snap-TF
algorithm, we also perform simulations for a large-scale sparse MIMO
array which is configured as a 15-element MRA with $M=15$, $N=79$,
and index set $\bm{\Omega}=\{1,2,3,6,11,16,27,38,49,60,66,72,78,79\}$,
corresponding to a 79-element virtual ULA \cite{schwartau2021large}.
The number of signal snapshots is set as $T=50$, and the number of
radio frequency (RF) chains is $N_{\text{RF}}=M$, if not otherwise
specified. We will also study the generalization performance to both
a smaller and a larger number of snapshots later. We consider the
general case of random received signal power. For each realization,
the path losses are generated as i.i.d. preliminary samples $\tilde{p}_{k}\sim\mathcal{U}(p_{\min},p_{\max})$
with $p_{\max}/p_{\min}\le10$, and then rescaled to enforce unit
mean power per source via $p_{k}\leftarrow K\,\tilde{p}_{k}/\sum_{j=1}^{K}\tilde{p}_{j}$,
which ensures $\frac{\max_{k}p_{k}}{\min_{k}p_{k}}\le10$. The average
received signal-to-noise ratio (SNR) is $\text{SNR}\triangleq10\log_{10}\!\big(\frac{\frac{1}{K}\sum_{j=1}^{K}p_{k}}{\eta}\big)$
dB. During the training stage, we generate a mixture of received signals
with SNR levels sampled uniformly from $[-20,20]$ dB and set $\ensuremath{\eta}$
accordingly to match the target SNR in each realization. For the Gaussian
signal case, the signals are generated with $s_{k}(t)\sim\mathcal{CN}(0,1)$,
while for the non-Gaussian data symbols, we mainly consider 16QAM
symbols, which are drawn randomly from alphabet sets $\mathcal{S}_{\text{16QAM}}=\{a+jb\mid a,b\in\{\pm\frac{1}{\sqrt{10}},\pm\frac{3}{\sqrt{10}}\}\}$.
The sources signals are mutually independent and also uncorrelated
with the noise if not otherwise stated, except for Section \ref{subsec:Coherence-Sources-Arising}
where we study the special case of coherence sources. For a given
$K$, we generate $\{\theta_{k}\}_{k=1}^{K}$ uniformly over $[\theta_{\min},\theta_{\max}]$
with a minimum angle separation constraint $\Delta_{\bm{\theta},\text{min}}\triangleq\min_{i\ne j}\left|\theta_{i}-\theta_{j}\right|,\forall i,j\in\{1,2,\ldots,K\}$.
This is realized by using the Poisson disk sampling \cite{bridson2007fast},
which yields an approximately uniform set over the interval while
enforcing a minimum angle separation constraint $\Delta_{\bm{\theta},\text{min}}$.
In the simulations, we set $\theta_{\min}=\tfrac{\pi}{6}$, $\theta_{\max}=\tfrac{5\pi}{6}$,
and $\Delta_{\bm{\theta},\text{min}}=\tfrac{\pi}{60}$, if not otherwise
stated. This relatively small minimum angle separation corresponds
to a practical and challenging scenario. The wavelength is set as
$\lambda=0.02$ meter, corresponding to 15 GHz carrier frequency in
the upper mid-band spectrum \cite{bazzi2025upper}.

\subsubsection{Neural Network Settings}

For the benchmark algorithms (DCR-T \cite{wu2022gridless}, DCR-G-Fro
\cite{barthelme2021DoA}, DCR-G-Aff \cite{barthelme2021DoA}, and
SRL-GD \cite{chen2025subspace}), we employ the same wide residual
network (WRN) as introduced in \cite{chen2025subspace} with 16 layers
and a widening factor of 8, i.e., the WRN-16-8 model, implemented
by the authors' open-source toolbox. All benchmarks adopt the same
backbone architecture. Their differences mainly lie in the loss function
and the output layer, which is an affine mapping with its output dimension
customized to each approach, as introduced in Section \ref{sec:Prior-Art}.
The input to the network is the real-valued representation of the
sample SCM of the SLA $\tilde{\mathbf{R}}_{\bm{\Omega}}\in\mathbb{C}^{M\times M}$,
obtained by concatenating the real and imaginary parts, with a shape
of $\mathbb{R}^{2\times M\times M}$. The output of the neural network
is the real-valued representation of the reconstructed SCM of the
virtual ULA $\hat{\mathbf{R}}\in\mathbb{C}^{N\times N}$, with a shape
of $\mathbb{R}^{2\times N\times N}$. Same as \cite{chen2025subspace},
the reconstructed SCM and the number of sources $K$ are then fed
into the Root-MUSIC algorithm to obtain the estimated DOAs. All the
benchmarks, as well as our proposed algorithm, use the number of sources
$K$ as an input, which is available at the BS according to 3GPP standards,
as explained before in footnote \ref{fn:In-3GPP-NR}. 
\begin{figure}[t]
\begin{centering}
\subfloat[]{
\centering{}\includegraphics[width=0.23\textwidth]{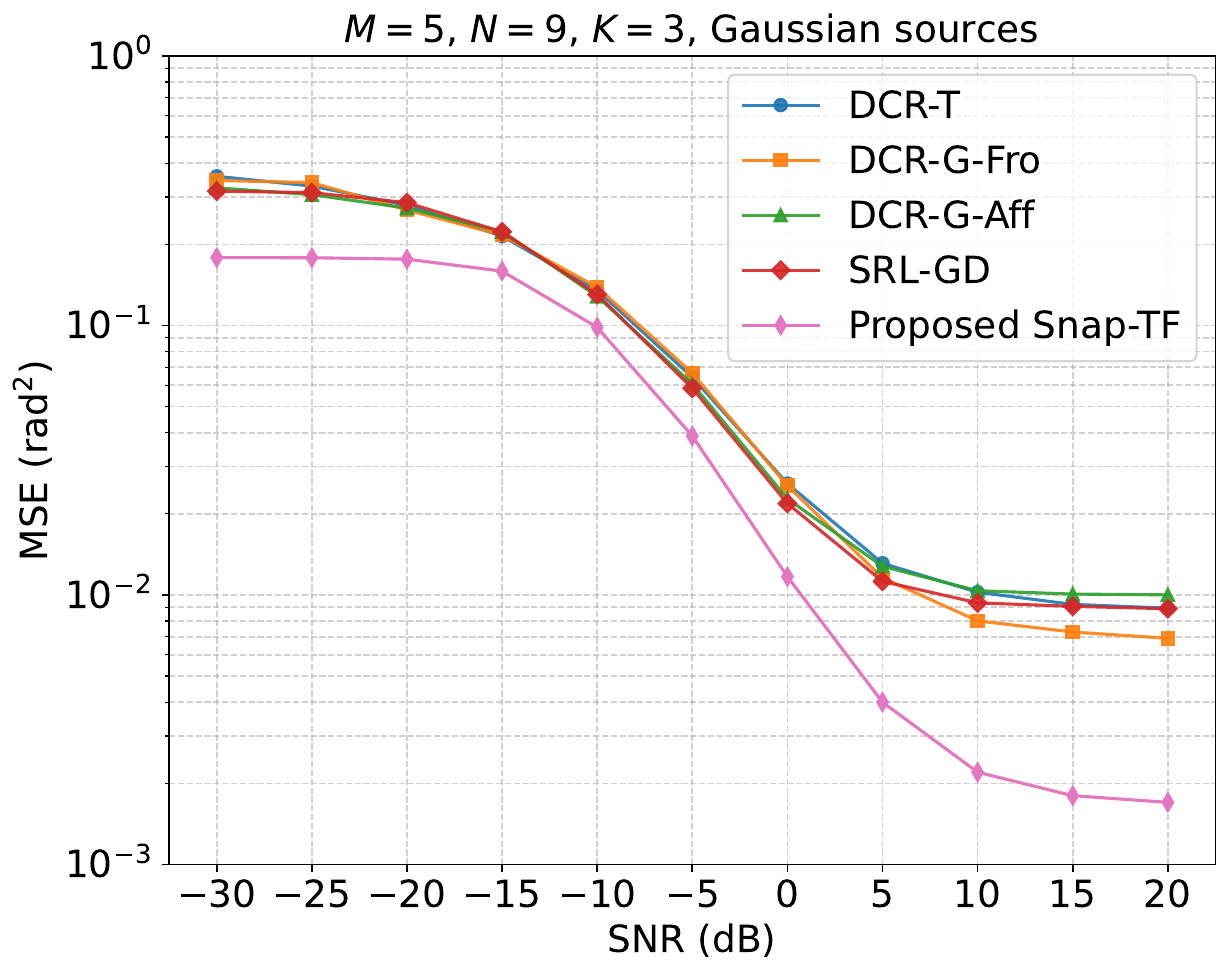}}\,\subfloat[]{\centering{}\includegraphics[width=0.23\textwidth]{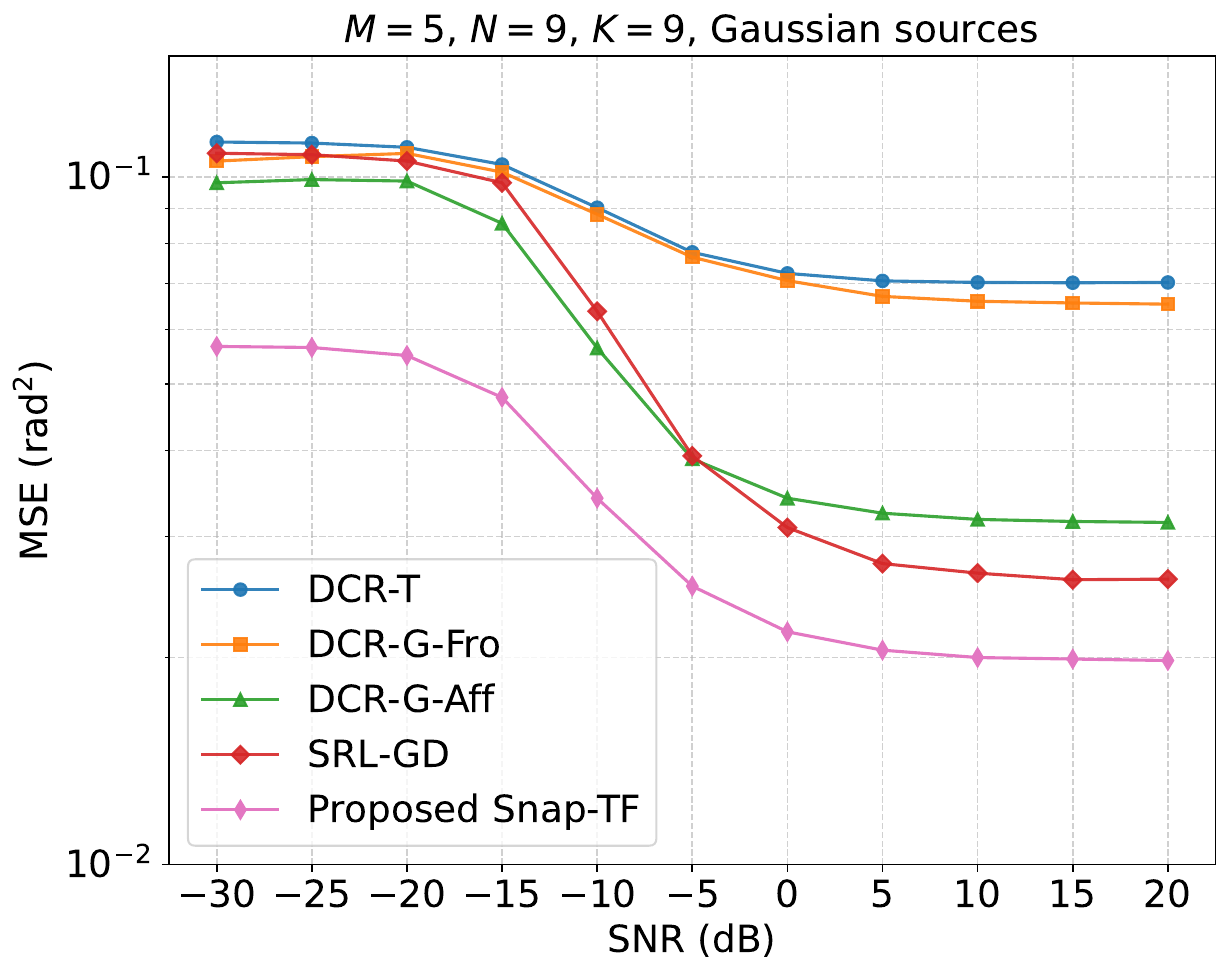}}
\par\end{centering}
\centering{}\subfloat[]{
\centering{}\includegraphics[width=0.23\textwidth]{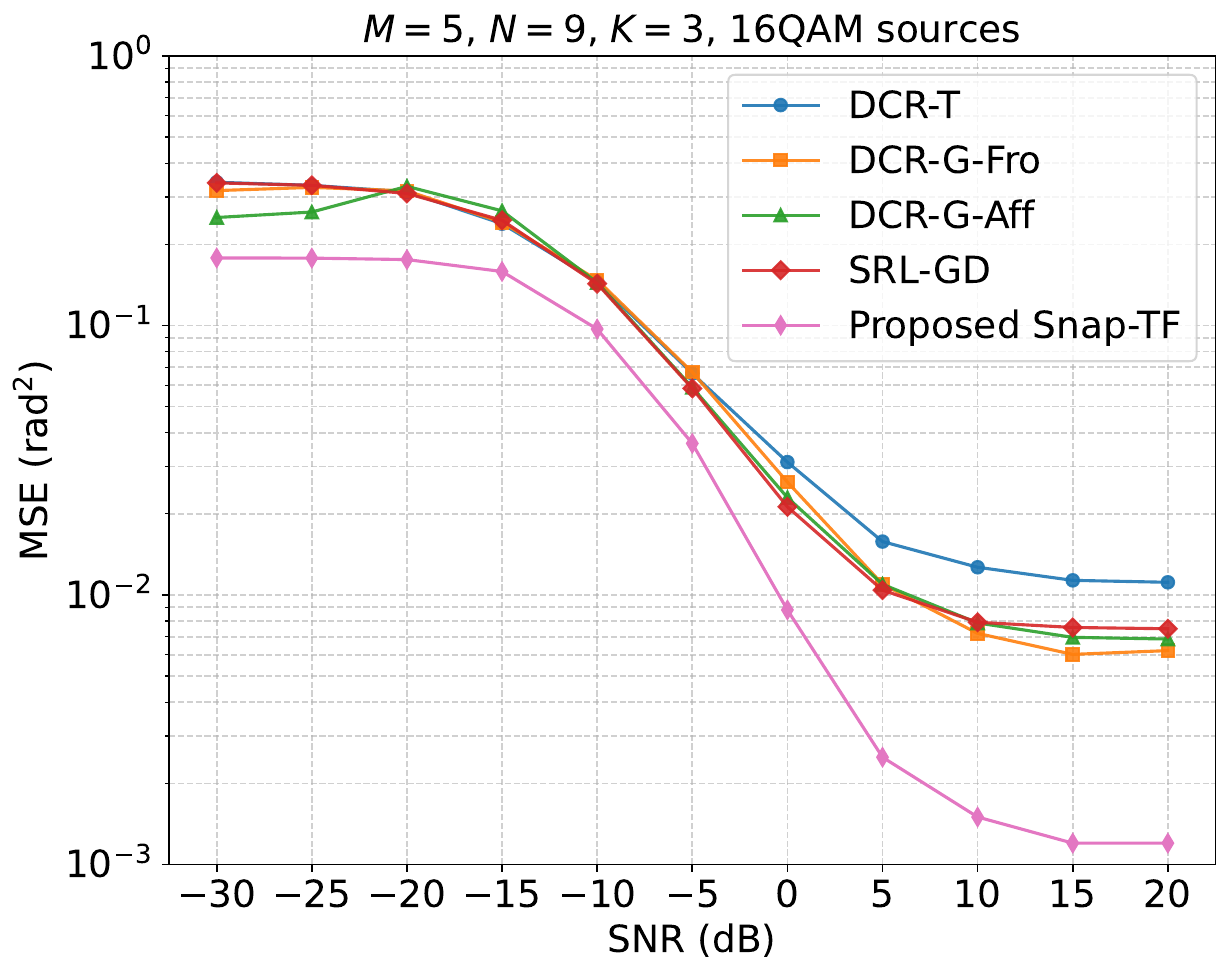}}\,\subfloat[]{
\centering{}\includegraphics[width=0.23\textwidth]{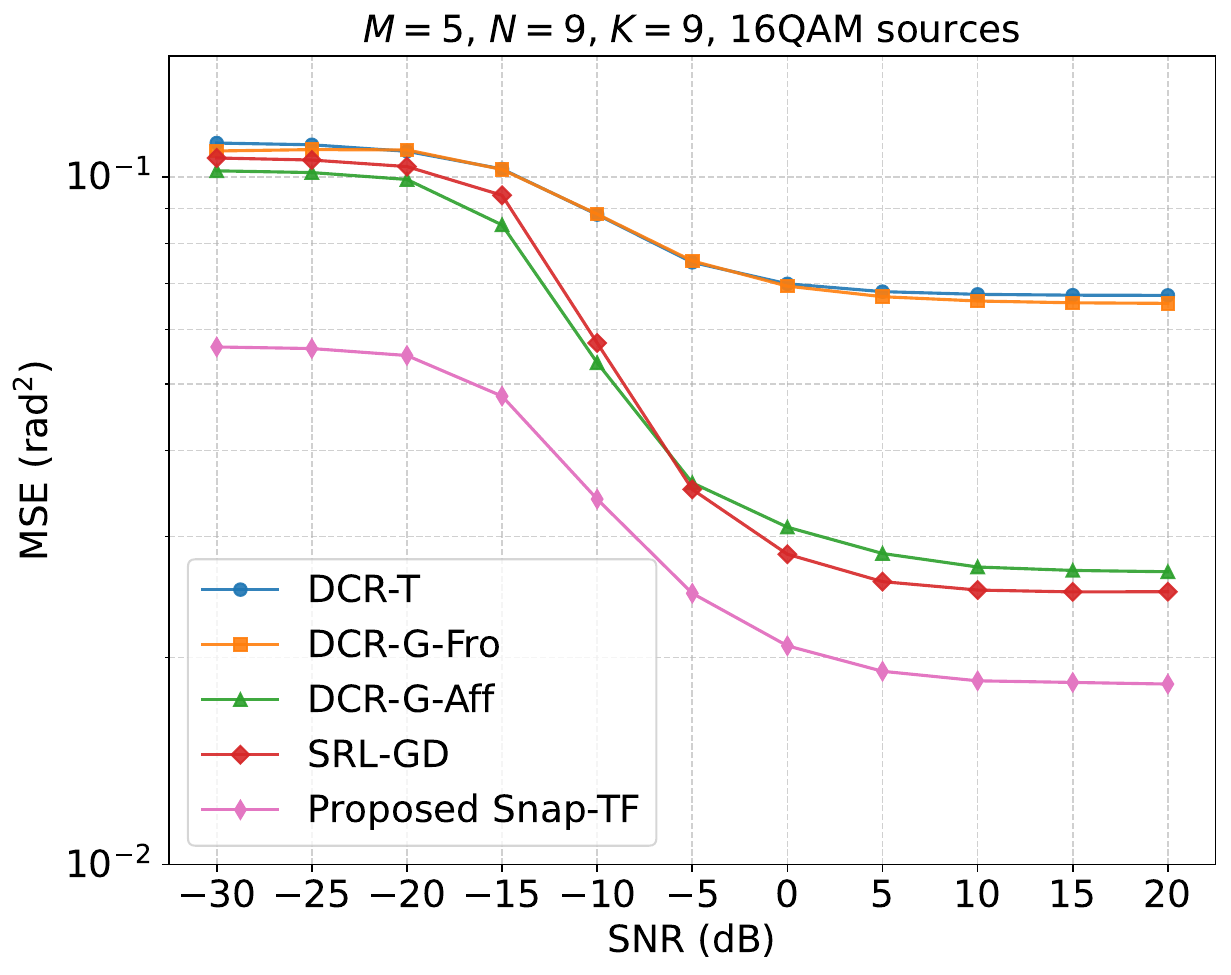}}\caption{MSE as a function of SNR under both Gaussian and non-Gaussian signals
with $M=5$ and $N=9$. The number of sources and signal types are
listed in the title of each subfigure. \label{fig:MSE-vs-SNR}}
\end{figure}

For the proposed Snap-TF, we set the attention activation dimension
as $d_{\text{attn}}=96$, and adopt $L=2$ transformer layers for
the Gaussian signals and $L=3$ layers in the case of non-Gaussian
modulated symbols. After the transformer layers, we apply an affine
mapping with a hidden dimension of 200 to transform the features to
the estimated DOAs. Specifically, for the proposed Snap-TF method,
we generate the DOA label by first sorting the $K$ angles in ascending
order, and then zero-pad the rest $(K_{\text{max}}-K)$ elements.
For example, let $K_{\text{max}}=9$ and $K=3$ with actual DOAs $\theta_{1}=\frac{\pi}{6}$,
$\theta_{2}=\frac{\pi}{3}$, and $\theta_{3}=\frac{\pi}{2}$, the
DOA label will be generated as $[\frac{\pi}{6},\frac{\pi}{3},\frac{\pi}{2},0,0,0,0,0,0]$.

\subsubsection{Training Settings}

We train the neural networks by using stochastic gradient descent
(SGD) with a batch size of 4096. The models are trained for 100 epochs
with the one-cycle scheduler of learning rates. For the benchmarks,
we use the same learning rates as suggested in \cite{chen2025subspace}
(0.05 for DCR-T, 0.01 for DCR-G-Fro, 0.005 for DCR-G-Aff, and 0.1
for SRL-GD). For the proposed Snap-TF algorithm, we utilize a learning
rate of 0.001. For each $K\in\{1,2,\ldots,K_{\text{max}}\}$, we generate
$2\times10^{6}$ training, $6\times10^{4}$ validation, and $1\times10^{4}$
testing samples, respectively. For $K_{\text{max}}=9$, this leads
to a total size of $9\times2\times10^{6}$, $9\times6\times10^{4}$,
and $9\times1\times10^{4}$ samples for training, validation, and
testing, respectively. We use PyTorch to train and test all the models
on one Nvidia A40 GPU.

\subsection{Performance\label{subsec:MSE-under-Gaussian}}

In Fig. \ref{fig:MSE-vs-SNR}, we plot the MSE of DOA estimation versus
SNR for $M=5$ and $N=9$ under both Gaussian (the top panel) and
non-Gaussian modulated sources (the bottom panel) with the number
of sources $K=3,9$ arranged from left to right. These respectively
represent cases where the number of sources is smaller and larger
than the number of antennas. The model is trained over a mixture of
SNR levels drawn uniformly from $[-20,20]$ dB, while tested in an
SNR range of $[-30,20]$ dB.

First, the MSE of all methods decreases almost monotonically as the
SNR increases. Our proposed Snap-TF algorithm always achieves the
lowest error across the entire SNR range. Moreover, the performance
gain is larger in the low-SNR regime. Second, comparing Gaussian versus
non-Gaussian modulated (16QAM) signals, i.e., top vs. bottom in each
column, the gap between Snap-TF and the benchmarks is larger for modulated
sources. Snap-TF can leverage higher-order statistics of modulated
symbols to localize better than covariance-based methods. These results
show that the proposed Snap-TF algorithm can localize the uplink DOAs
solely based on the unknown uplink data symbols, which is performed
at the same time with data transmission, without additional overhead.

Third, increasing the number of sources from $K=3$ to $K=9$ raises
the MSE for all methods, since $K=3$ (smaller than $M-1$) falls
within the physical array's degree of freedom while $K=9$ (larger
than $M-1$) require exploiting the virtual array to resolve the sources.
In both cases, Snap-TF consistently delivers better DOA estimation
accuracy than all the benchmark methods. This extended localization
capability is particularly useful for dense 6G networks with more
sources than antennas (e.g., low-altitude UAV swarms) \cite{lin2025tale}.

\subsection{Generalization Capability\label{subsec:Generalization-Capability}}

\begin{figure}[t]
\begin{centering}
\subfloat[]{
\centering{}\includegraphics[width=0.23\textwidth]{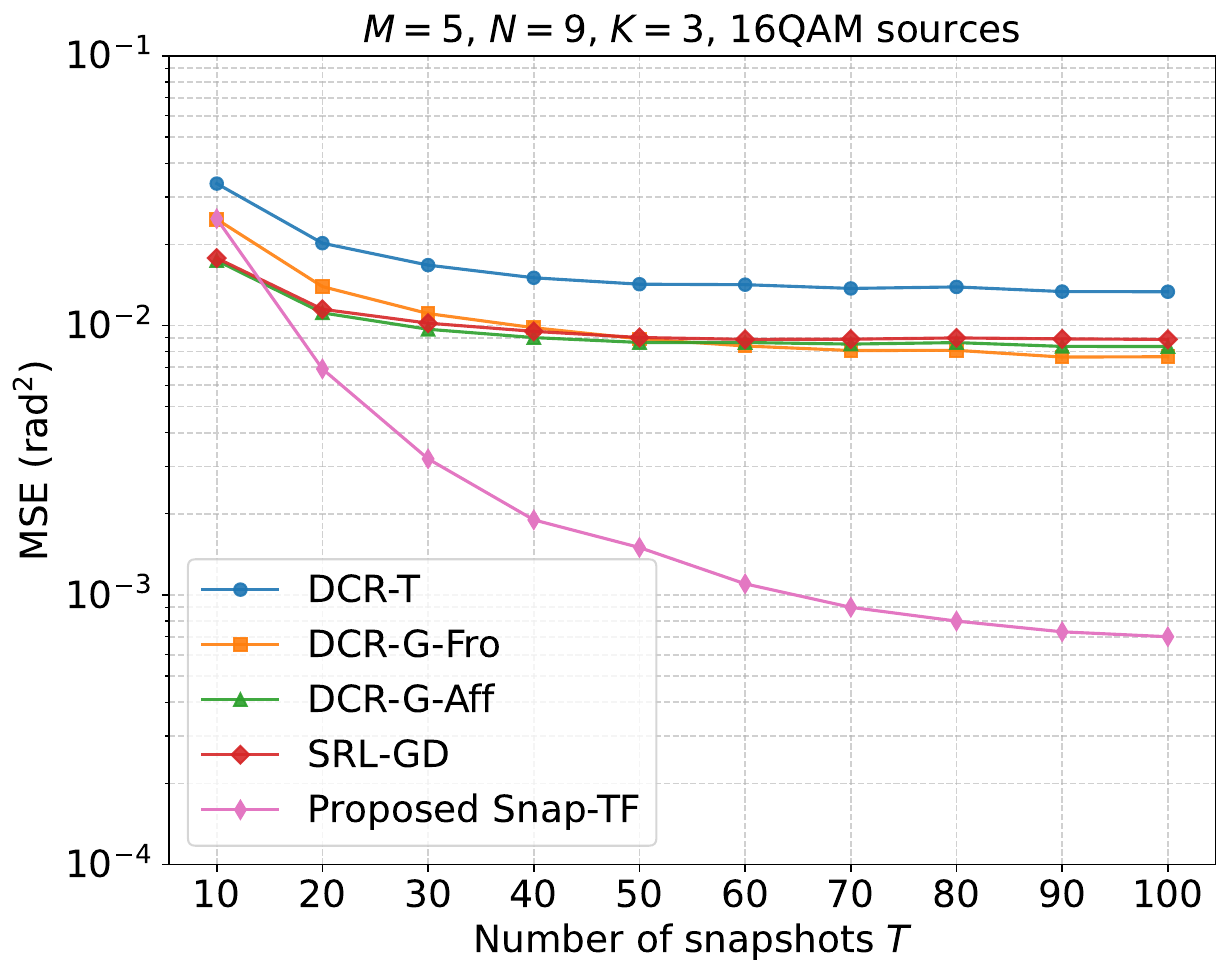}}\,\subfloat[]{\centering{}\includegraphics[width=0.23\textwidth]{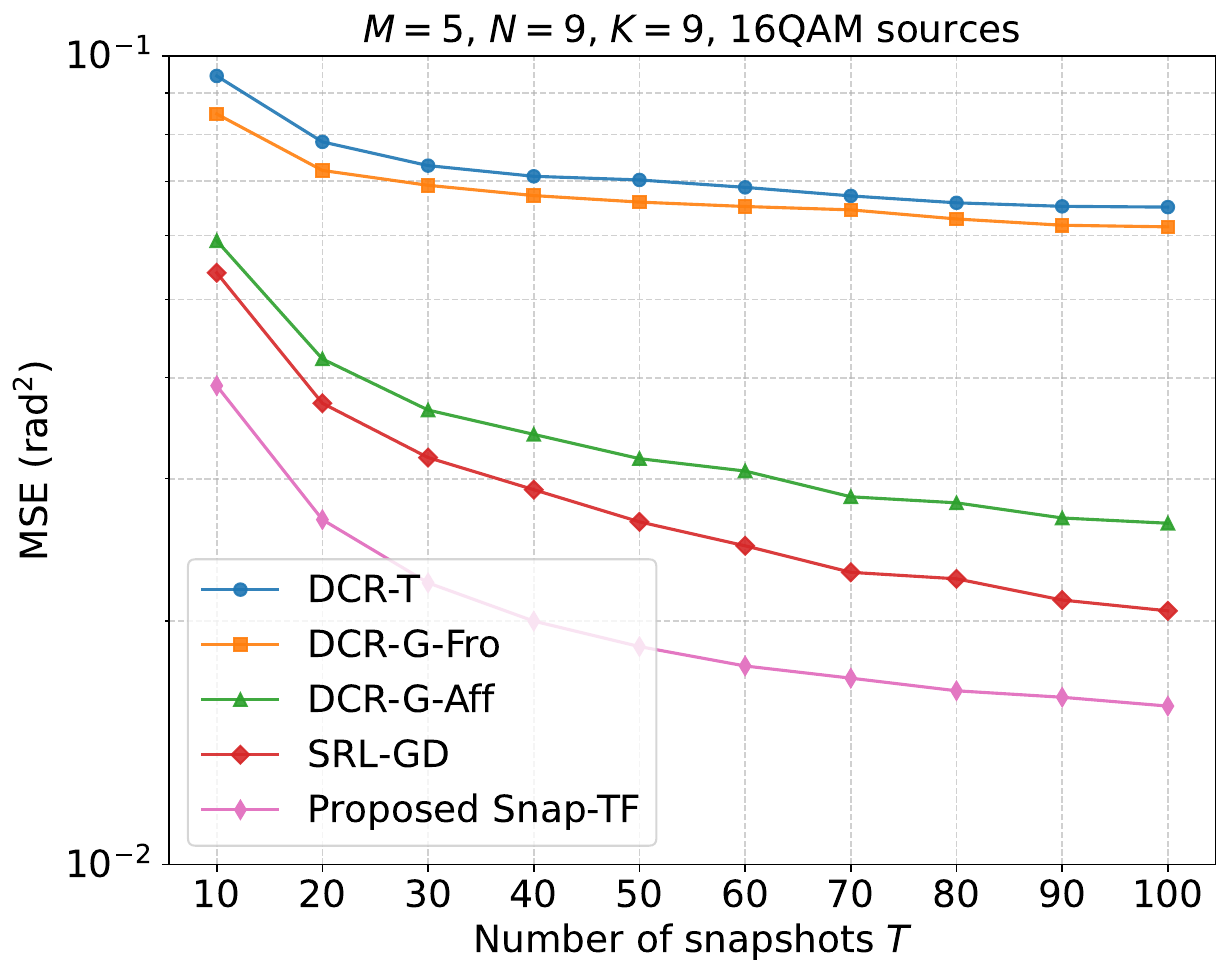}}
\par\end{centering}
\begin{centering}
\subfloat[]{\centering{}\includegraphics[width=0.23\textwidth]{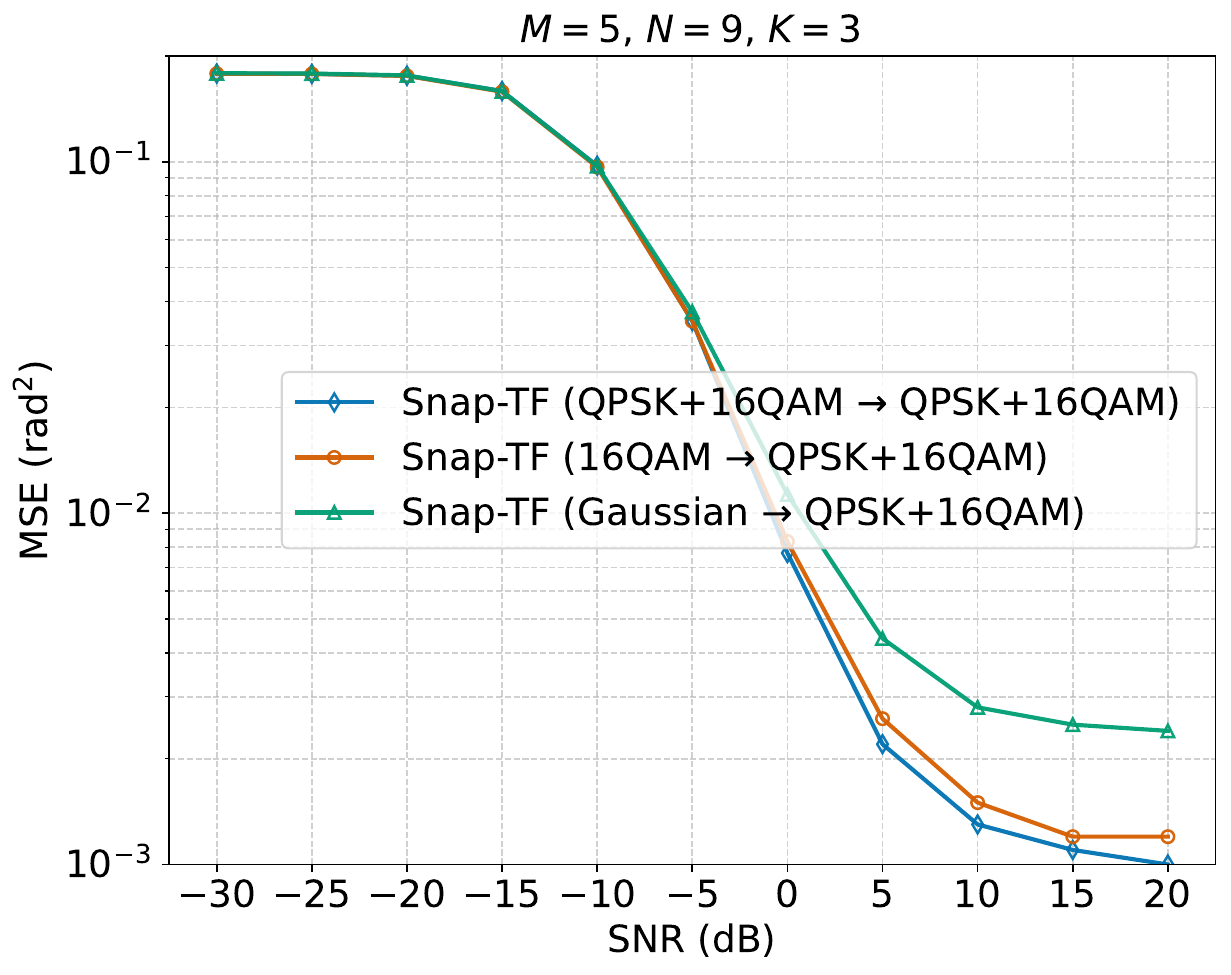}}\,\subfloat[]{\centering{}\includegraphics[width=0.23\textwidth]{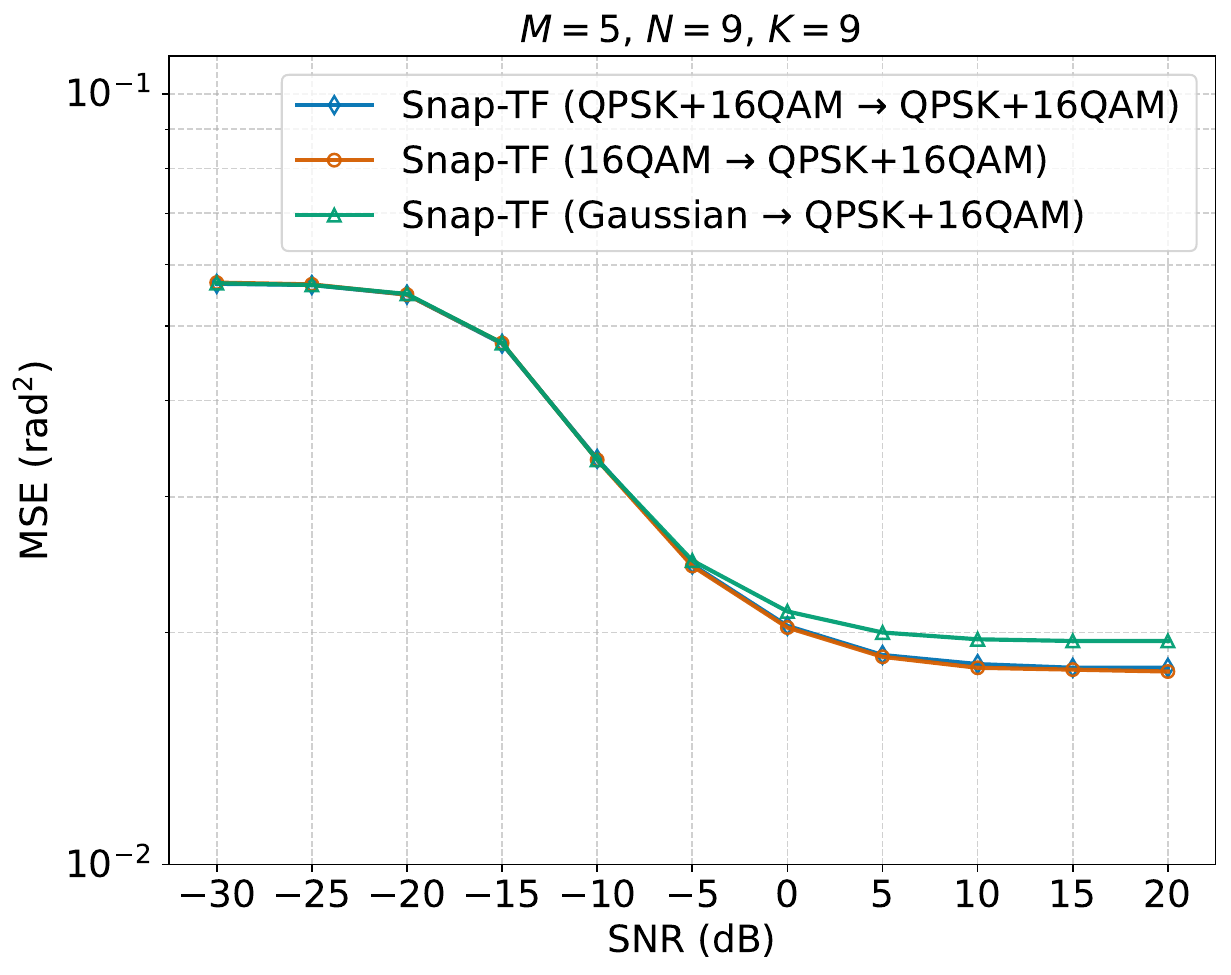}}
\par\end{centering}
\centering{}\caption{Generalization performance to different numbers of snapshots $T$
and different modulation types. The model is trained under $T=50$
and $\Delta_{\bm{\theta},\min}=\frac{\pi}{60}$, and directly tested
in all the different settings. \label{fig:Generalization-performance-to}}
\end{figure}

In Fig. \ref{fig:Generalization-performance-to}, we show the generalization
capability of the proposed Snap-TF algorithms to practical mismatches,
including the number of snapshots $T$ and the different modulation
types. All the methods shown in the figure are trained under $T=50$
and $\Delta_{\bm{\theta},\min}=\frac{\pi}{60}$, and directly tested
in different cases without retraining or tuning. During testing, the
SNR is set as 10 dB. 

In the top panel, we evaluate generalization across the numbers of
snapshots $T$ from 10 to 100 while fixing the minimum angle separation
at $\Delta_{\bm{\theta},\min}=\frac{\pi}{60}$. Despite being trained
at $T=50$, Snap-TF achieves superior performance for $20\le T\le100$,
demonstrating adaptability to varying numbers of snapshots while maintaining
robustness. A notable exception is $T=10$ in Fig. \ref{fig:Generalization-performance-to}(a),
where Snap-TF slightly underperforms the best benchmark. We conjecture
that, because $T=10$ is very small, the sample estimates of higher-order
statistics exploited by Snap-TF have high variance. Moreover, this
test uses $K<M$ (fewer sources than antennas), a regime in which
second-order statistics (used by the benchmarks) already perform well.
As a result, higher-order information could be misleading at such
small $T$, leading to a disadvantage for Snap-TF at $T=10$. The
slight fluctuations are attributed to the fact that these are out-of-distribution
cases that the models were not trained on.

In the bottom panel, we evaluate the generalization performance of
Snap-TF to different, or even mixed, modulation types that it was
not trained on. Specifically, we set up a ``mixed modulation type''
case where each source transmits in either QPSK or 16QAM modulation.
We draw a Bernoulli distributed mask, i.e., $\mathrm{Bernoulli}(0.5)$,
for every symbol transmitted by each source in each snapshot. If the
random mask is 1, the source transmits a QPSK symbol. Otherwise, if
the mask equals 0, the source transmits a 16QAM symbol instead. We
test three variants of the proposed Snap-TF algorithm on such a ``mixed
modulation type'' case. The left-hand-side of the arrow denotes the
dataset where the model is trained on, while the right-hand-side denotes
the testing dataset of the model. For example, 16QAM $\rightarrow$
QPSK+16QAM denotes the case where a model trained on 16QAM sources,
but tested on mixed modulation types of QPSK+16QAM sources. Simulation
results show that under both $K=3$ and $K=9$ sources, the curve
trained on 16QAM only is very close to the model trained on the mixed
QPSK+16QAM data. This suggests that after trained on a single non-Gaussian,
discrete constellation with amplitude variation, Snap-TF has the potential
to generalize to different, and even mixed, modulation types. In contrast,
the Gaussian $\rightarrow$ QPSK+16QAM model shows a notable performance
degradation at medium to high SNR, which is likely attributed to the
absence of higher-order statistics when the model is trained on Gaussian
signals.

\subsection{Scalability\label{subsec:Scalability-to-Large-Scale}}

\begin{figure}[t]
\centering{}\subfloat[]{\centering{}\includegraphics[width=0.23\textwidth]{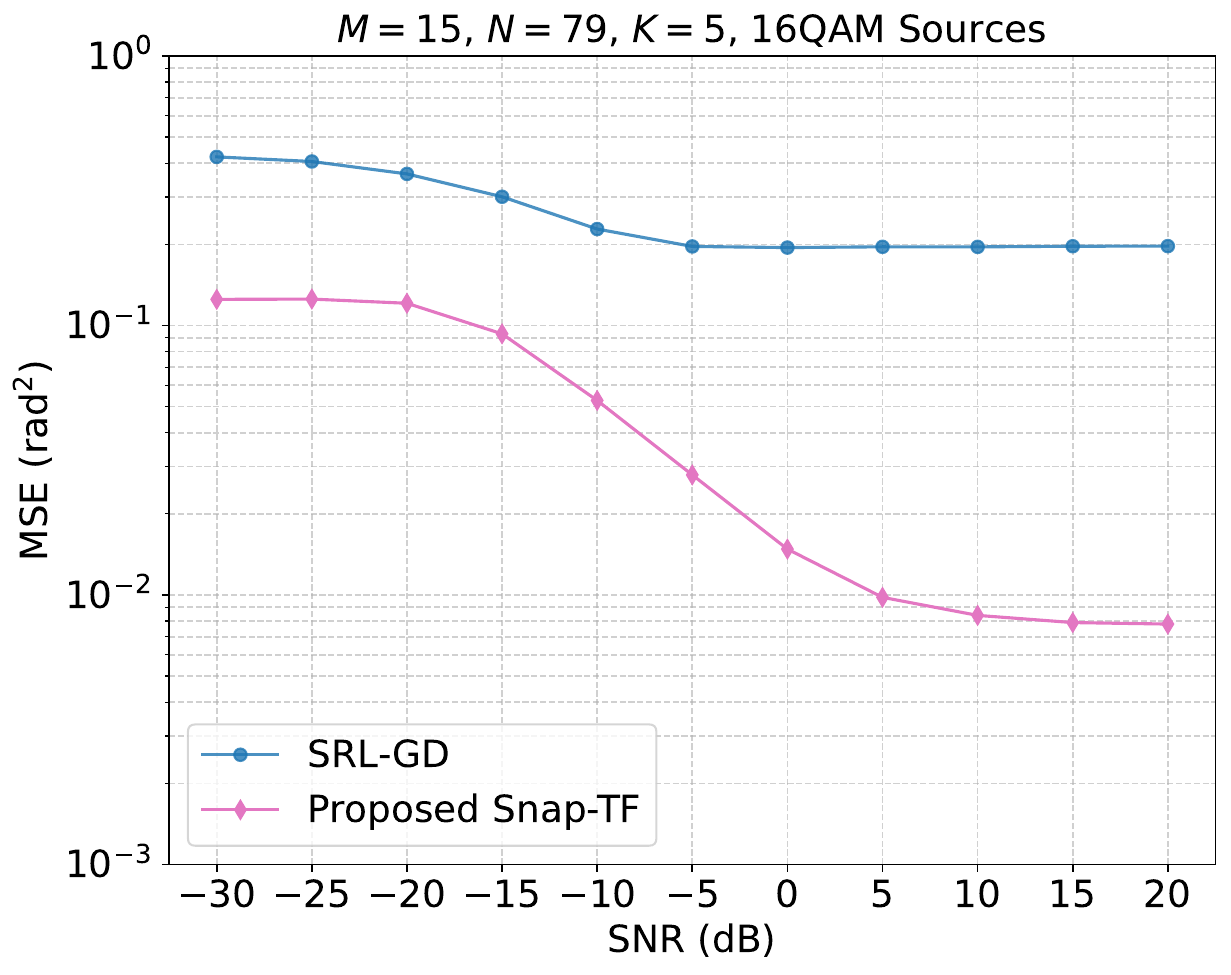}}\,\subfloat[]{\centering{}\includegraphics[width=0.23\textwidth]{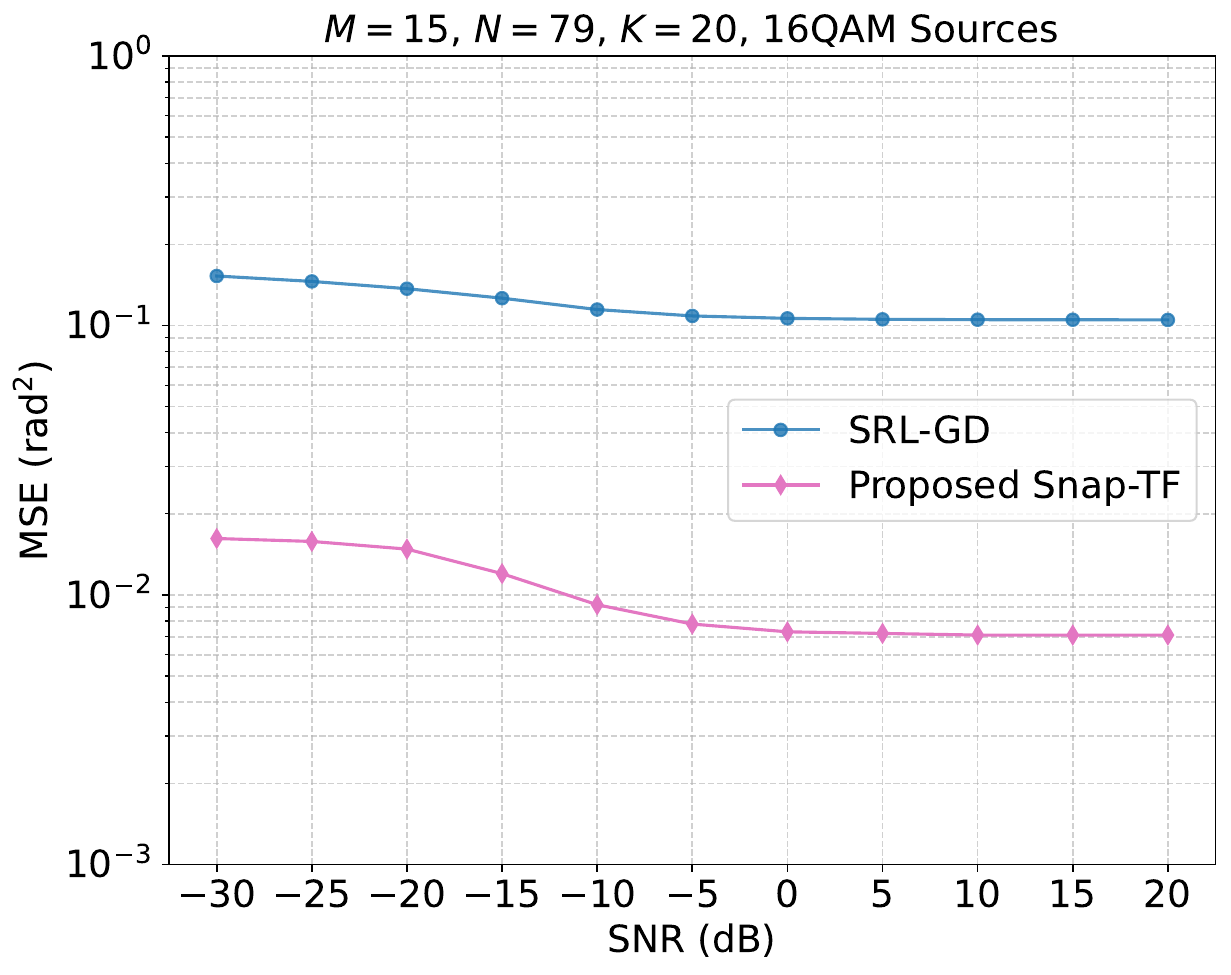}}\caption{Scalability of the proposed Snap-TF algorithm in large-scale systems.
\label{fig:Scalability}}
\end{figure}

In Fig. \ref{fig:Scalability}, we plot MSE-SNR curves to show that
the proposed Snap-TF algorithm is scalable for large-scale MIMO systems.
In this set of experiments, we adopt a 15-element MRA with $M=15$,
which corresponds to a virtual ULA of $N=79$ elements, as introduced
in Section \ref{subsec:System-Settings}. We consider the cases where
the numbers of sources are $K=5$ and $K=20$, respectively. Other
system settings are the same as described in Section \ref{subsec:System-Settings}.
In this case, the benchmark methods will suffer from computational
issues. The Root-MUSIC algorithm in the inference process requires
the eigen-decomposition of an $N\times N$ complex-valued matrix,
which is challenging to compute. In addition, the training process
of the benchmarks DCR-G-Aff and SRL-GD both requires matrix inversion
or decomposition to compute the loss function, which also undermines
the scalability of their training process. In view of the complexity
issue, we only compare with the most competitive benchmark, i.e.,
the SRL-GD algorithm, in the simulations. 

As shown, the proposed Snap-TF algorithm significantly outperforms
the benchmark both when the number of sources $K$ is below the number
of physical antennas ($K=5$) and when it exceeds it ($K=20$). The
performance margin is even larger compared to that in the small-scale
system. The results verify the scalability of the proposed algorithm
in large-scale systems, and demonstrate its capability to localize
more sources than antennas with unknown data symbols. The proposed
algorithm also has significant advantages in complexity in large-scale
systems, as will be detailed later.

\subsection{Complexity\label{subsec:Complexity}}

\begin{figure}[t]
\centering{}\subfloat[]{\centering{}\includegraphics[width=0.23\textwidth]{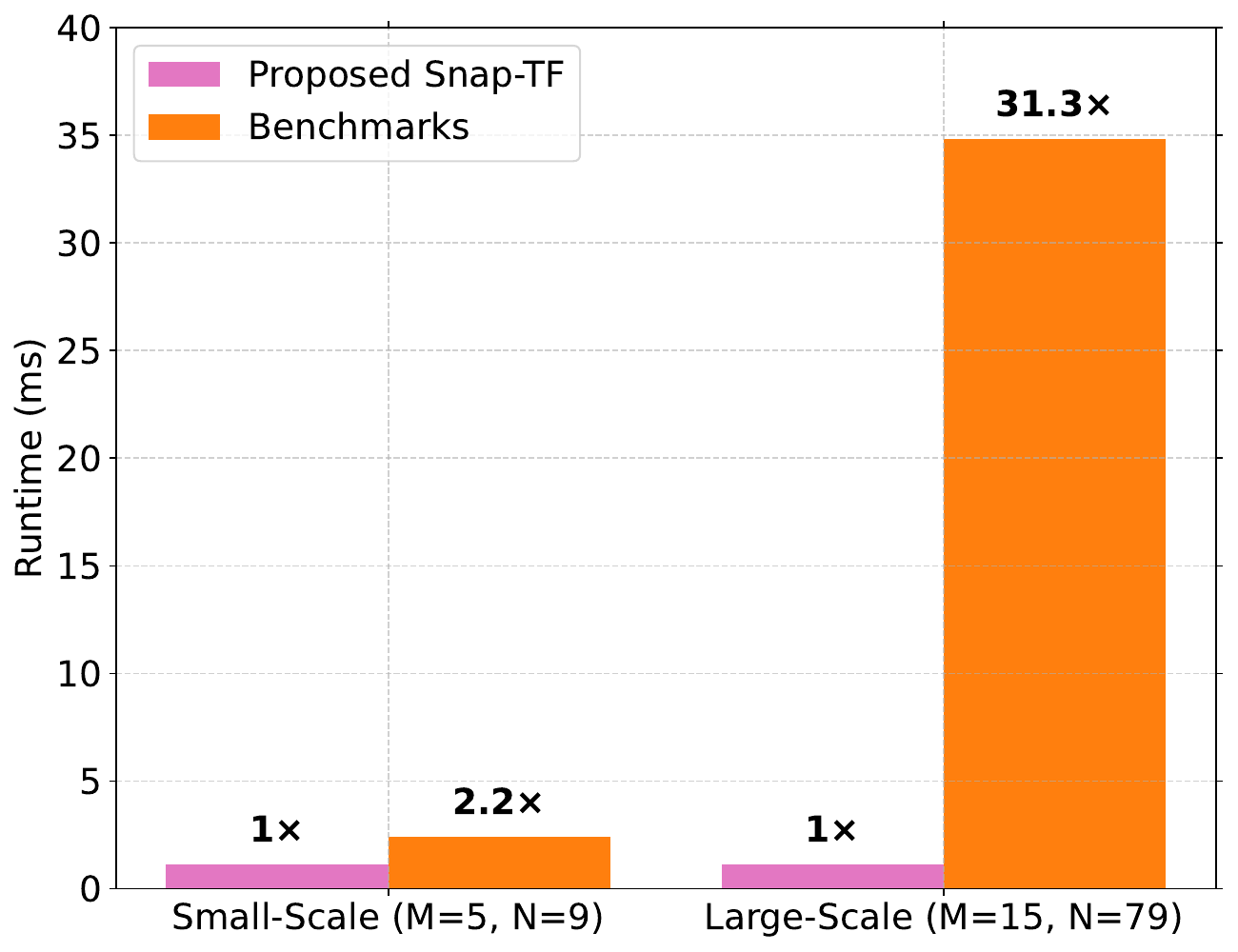}}\,\subfloat[]{\centering{}\includegraphics[width=0.23\textwidth]{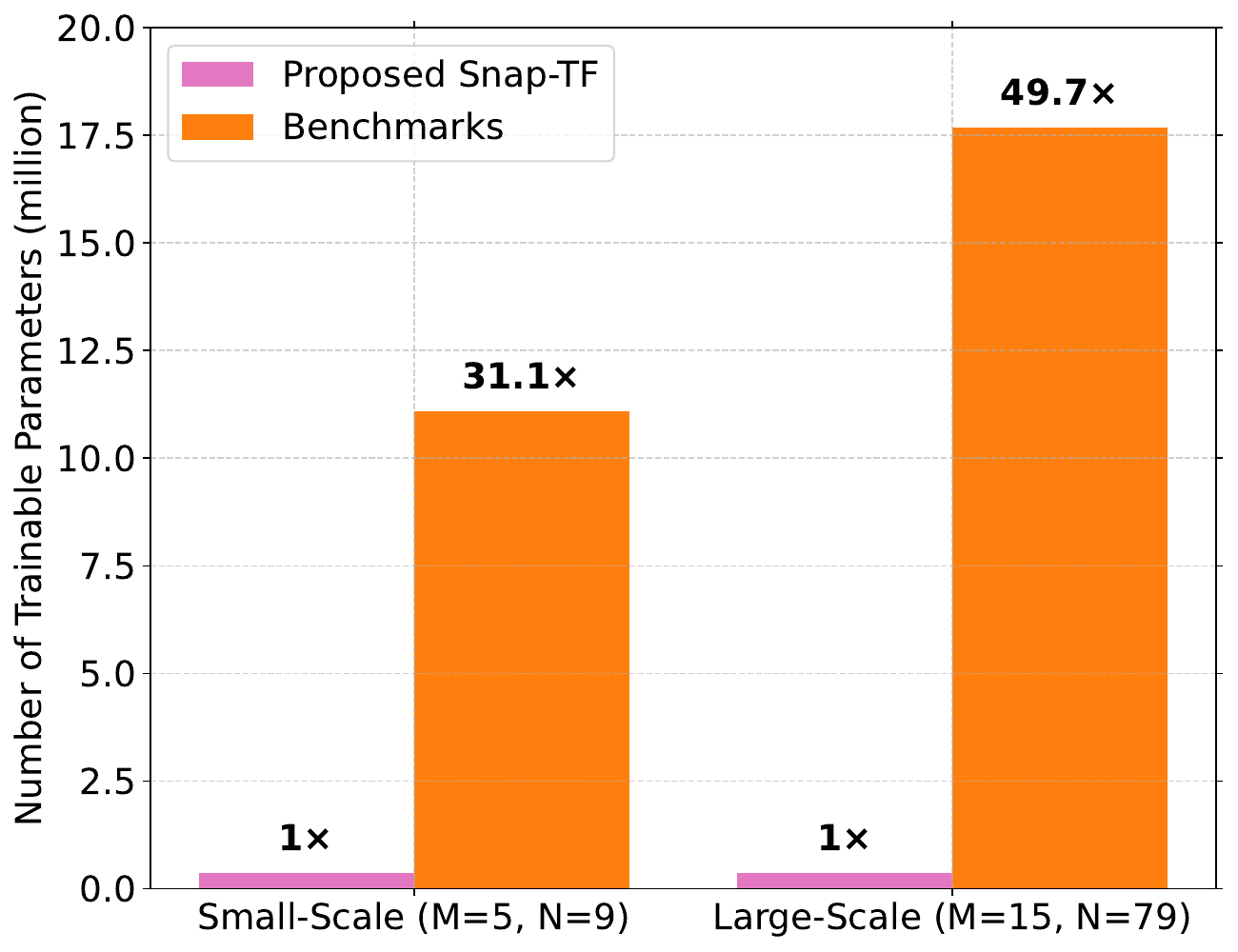}}\caption{Comparison of complexity, including (a) runtime and (b) number of
trainable parameters, under different system scales. \label{fig:Comparison-of-runtime}}
\end{figure}

In Fig. \ref{fig:Comparison-of-runtime}, we compare the complexity
of different algorithms under various system scales including both
runtime and number of trainable parameters. The runtime and number
of parameters of the benchmarks are almost the same since they all
use the same WRN network as backbone followed by the Root-MUSIC algorithm.
We use SRL-GD as an indicator. The comparison is performed under 16QAM
symbols, in which case we adopt 3 layers for the proposed Snap-TF
algorithm. 

As seen in Fig. \ref{fig:Comparison-of-runtime}(a), the proposed
Snap-TF algorithm consistently achieves sub-millisecond level runtime
under both system configurations, approximately 1.103 ms for the small-scale
case ($M=5$, $N=9$) and 1.112 ms for the large-scale case ($M=15$,
$N=79$). By contrast, the benchmark DL-based subspace methods, which
rely on matrix decompositions, exhibit a $2.2\times$ slowdown over
Snap-TF in the small-scale case but suffer a dramatic $31.3\times$
increase in runtime when scaled up. This disparity underscores the
heavy cost of the benchmarks, including $O(N^{3})$ operations such
as matrix decomposition and inversion, making them computationally
prohibitive in large-scale MIMO systems. By contrast, Snap-TF circumvents
this bottleneck by eliminating the need for explicit matrix decomposition.
It maintains a moderate runtime even in large-scale setups, and is
suitable for real-time applications.

In Fig. \ref{fig:Comparison-of-runtime}(b), we compare the number
of trainable parameters. The number for the benchmark is around 11
million and 17 million in the small-scale and large-scale setups,
respectively, while that of the proposed Snap-TF algorithm (with 3
layers) is around 0.356 million in both setups. The proposed algorithm
is $31.1\times$ and $49.7\times$ more efficient in parameters, respectively,
which is lightweight for practical deployment.

\subsection{Ablation Studies\label{subsec:Ablation-Studies}}

\begin{figure}[t]
\centering{}\includegraphics[width=0.23\textwidth]{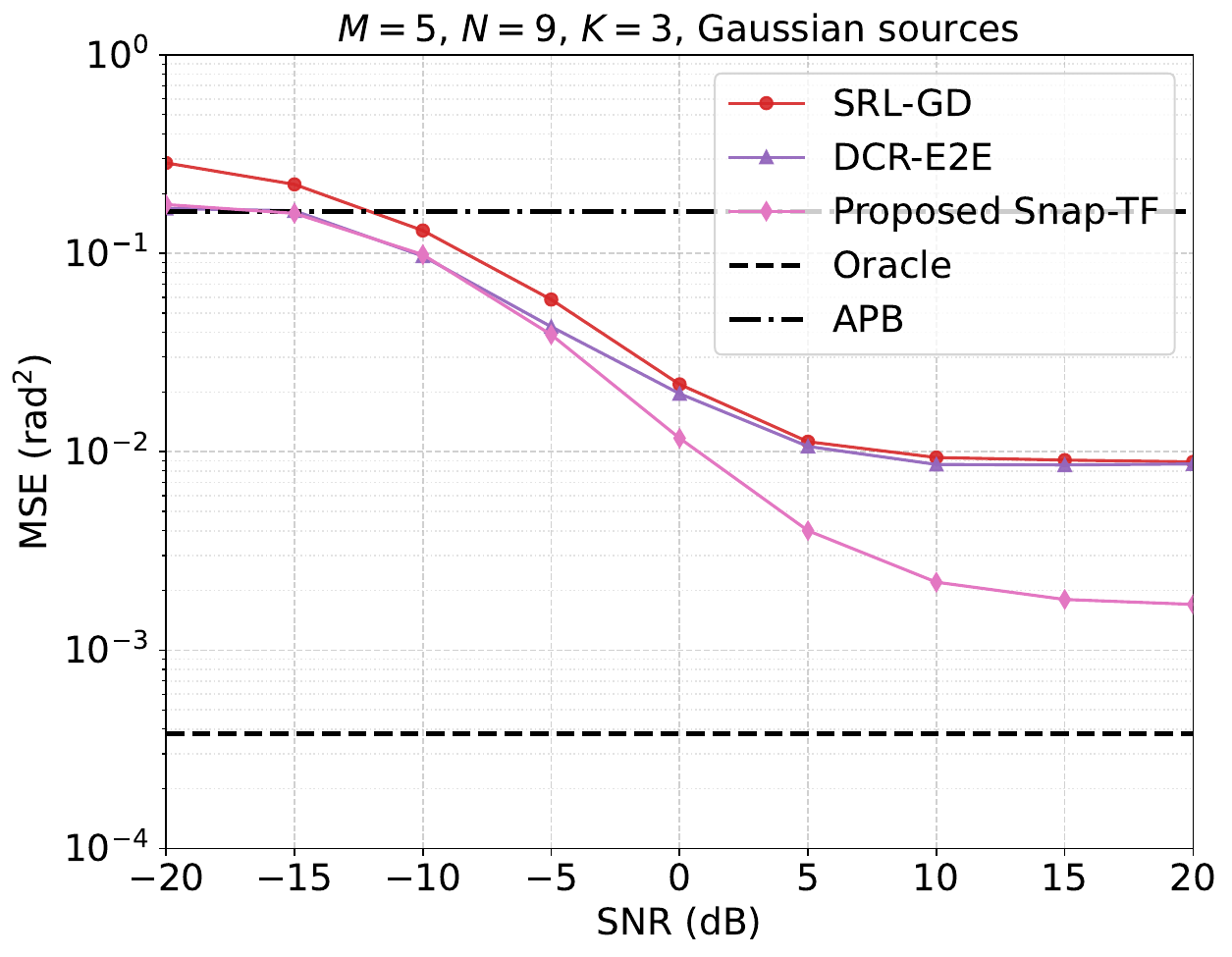}\caption{Ablation studies of the reasons behind the performance gains. \label{fig:Ablation-gains}}
\end{figure}

We first conduct an ablation study to identify the sources of performance
gains by isolating two key factors: (i) raw snapshot processing versus
sample SCM-based input, and (ii) end-to-end (E2E) training versus
two-stage algorithm (learning covariance reconstruction followed by
Root-MUSIC). We compare five methods in Fig. \ref{fig:Ablation-gains}:
the proposed Snap-TF (raw snapshots + E2E), SRL-GD \cite{chen2025subspace}
(sample SCM + two-stage), DCR-E2E (sample SCM + E2E), an Oracle bound\footnote{This is obtained by applying Root-MUSIC to the sample SCM of a physical
10-element ULA (matching the virtual aperture of our 5-element MRA)
under identical conditions ($\text{SNR}=20$ dB, $T=50$). This represents
an ideal case and may be loose as it uses physical 10-element ULA.}, and the \textit{a priori} bound (APB) \cite{Zhang2023Ziv} (performance
bound at low SNRs). 

The results shown in Fig. \ref{fig:Ablation-gains} reveal two distinct
advantages of the proposed Snap-TF algorithm. First, at high SNR,
both sample SCM-based methods (SRL-GD and DCR-E2E) plateau at $10^{-2}$
rad$^{2}$ regardless of training approach, demonstrating that the
sample SCM from $T=50$ snapshots creates a finite-sample bottleneck.
In contrast, Snap-TF achieves $2\times10^{-3}$ rad$^{2}$ by directly
processing raw snapshots through self-attention, which preserves the
information of each snapshot and adaptively weights interactions between
different snapshots. Second, at low SNR, both E2E methods (Snap-TF
and DCR-E2E) converge to the APB while SRL-GD remains notably worse,
showing that E2E training effectively lets neural networks learn DOA
priors. Thus, Snap-TF's superiority stems from both efficient raw
snapshot processing via self-attention (high-SNR advantage) and E2E
learning (low-SNR advantage). At the same time, both components are
more lightweight and efficient compared to prior arts and achieve
30$\times$ parameter reduction and faster inference as discussed
before.

\begin{figure}[t]
\begin{centering}
\subfloat[]{
\centering{}\includegraphics[width=0.23\textwidth]{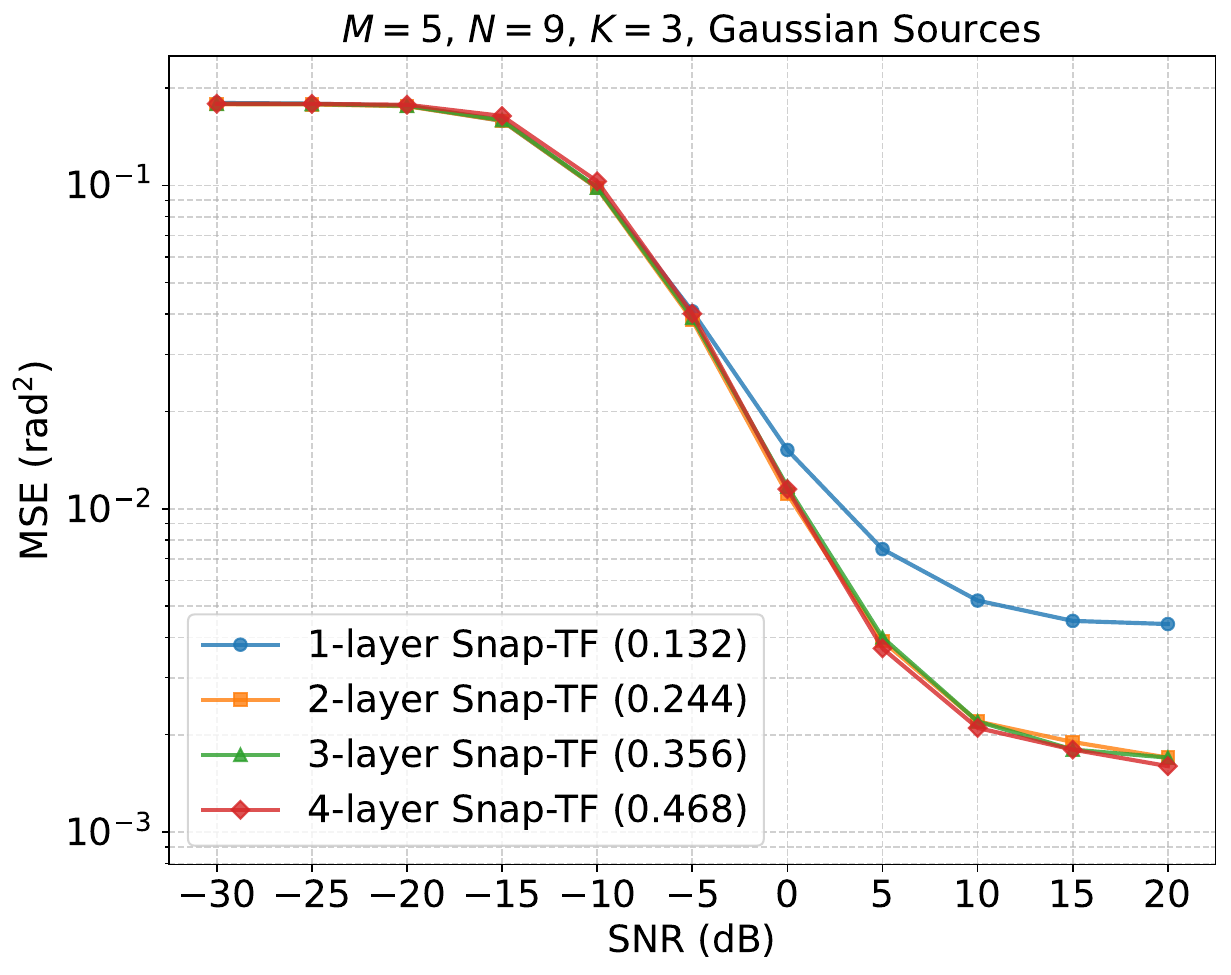}}\,\subfloat[]{\centering{}\includegraphics[width=0.23\textwidth]{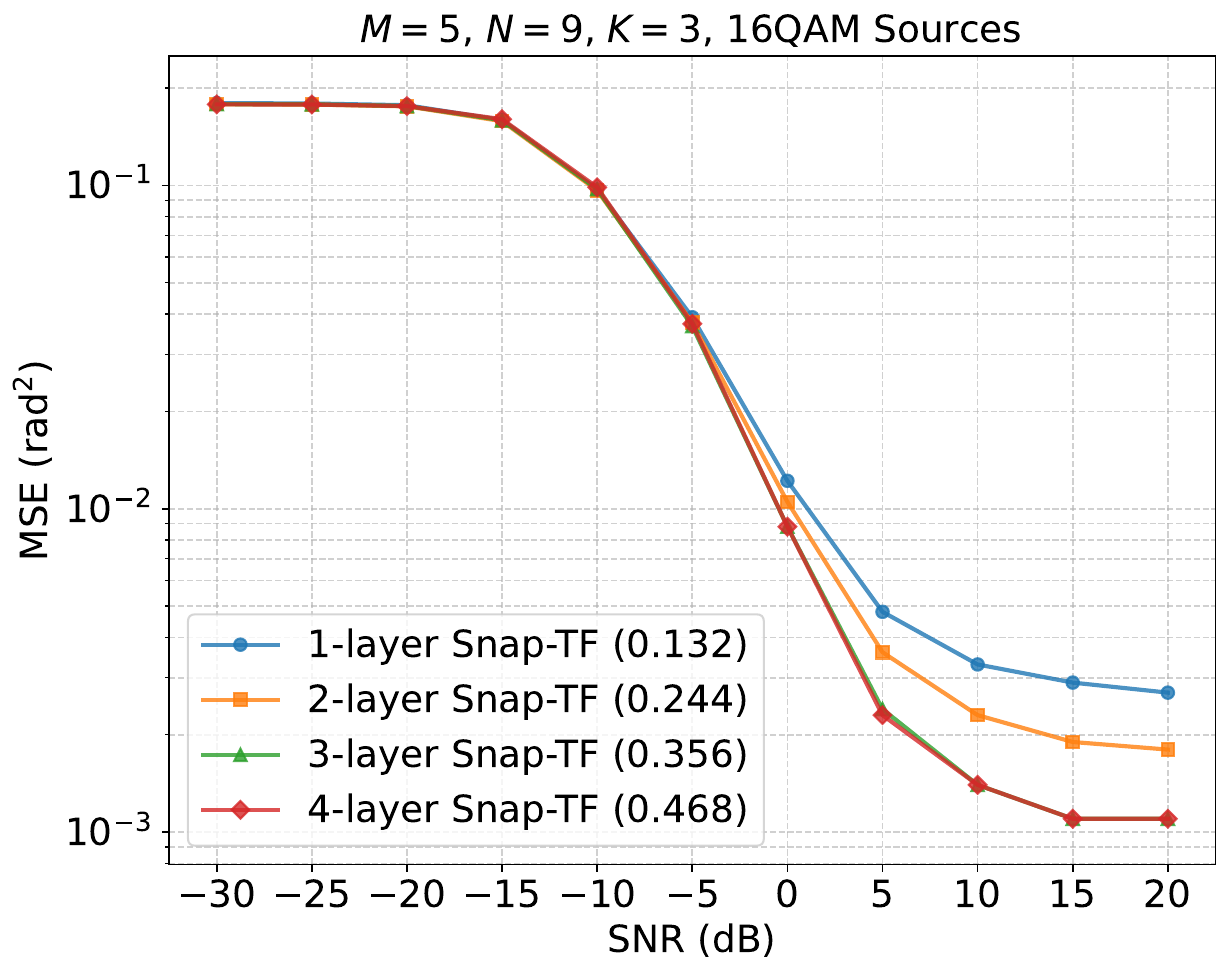}}
\par\end{centering}
\centering{}\caption{Ablation studies for the required layers of Snap-TF model. \label{fig:Ablation-studies}}
\end{figure}

We also conduct ablation studies to demonstrate Snap-TF's capability
to effectively learn and exploit higher-order statistical for DOA
estimation. In the case of Gaussian signals, $k=2$ (second) order
statistics are the sufficient statistics. As discussed in Section
\ref{sec:Proposed-Snap-TF}, only $L=\text{\ensuremath{\lceil\log_{2}(k+1)\rceil}}=2$
single-head transformer layers are sufficient. This is verified by
the simulation results in Fig. \ref{fig:Ablation-studies}(a), where
we observe the performance starts to saturate when the model uses
more than two layers. The saturation behavior provides evidence that
Snap-TF has successfully captured all the relevant statistical information
available in the Gaussian case, with additional layers offering diminishing
returns. Similarly, in the non-Gaussian case, 16QAM symbols have non-zero
$k=4$ (fourth) order cumulants \cite{yuen1997DOA}, and hence $L=\text{\ensuremath{\lceil\log_{2}(k+1)\rceil}}=3$
layers are required. The results in Fig. \ref{fig:Ablation-studies}(b)
verify our hypothesis, which show that the performance keeps improving
in the first three layers, and saturates thereafter. The performance
pattern matches our prediction, and confirms that Snap-TF effectively
captures the higher-order statistics required for non-Gaussian DOA
estimation. This also suggests a lightweight 3-layer model is sufficient
for multi-source localization in cellular systems based on unknown
uplink data symbols. We also label the number of parameters in the
unit of million in the legends. 

\subsection{Multipath Propagation and Coherent Sources \label{subsec:Coherence-Sources-Arising}}

We now discuss the impact of multipath propagation. As an example,
assume two signals are coherent with $s_{2}(t)=\alpha s_{1}(t)$ due
to multipath propagation where $\alpha\in\mathbb{C}$. Although the
system model in (\ref{eq:system-model}) still holds, the pair collapses
to a single second-order mode: the source vector becomes $\tilde{\mathbf{s}}(t)=[s_{1}(t),s_{3}(t),\ldots,s_{K}(t)]^{\mathsf{T}}$
and the array response matrix becomes $\tilde{\mathbf{A}}_{\bm{\Omega}}(\bm{\theta})=[\mathbf{a}_{\bm{\Omega}}(\theta_{1})+\alpha\mathbf{a}_{\bm{\Omega}}(\theta_{2}),\mathbf{a}_{\bm{\Omega}}(\theta_{3}),\ldots,\mathbf{a}_{\bm{\Omega}}(\theta_{K})]$,
so the coherent pair $\{s_{1}(t),s_{2}(t)\}$ appears as a single
source with ``combined'' steering vector $\mathbf{a}_{\bm{\Omega}}(\theta_{1})+\alpha\mathbf{a}_{\bm{\Omega}}(\theta_{2})$
that generally lies outside the original array response matrix $\mathbf{A}_{\bm{\Omega}}(\bm{\theta})$.
As a result, SCM-based MUSIC fails to resolve the pair. By contrast,
higher-order statistics (e.g., fourth-order) remain informative under
coherence sources because they include per-source self-products such
as $\mathbf{a}_{\bm{\Omega}}(\theta_{1})\otimes\mathbf{a}_{\bm{\Omega}}(\theta_{1})$
and $\mathbf{a}_{\bm{\Omega}}(\theta_{2})\otimes\mathbf{a}_{\bm{\Omega}}(\theta_{2})$
that do not collapse, thereby restoring separability \cite{yuen1997DOA}.
Our proposed snap-TF implicitly leverages such higher-order statistics
from non-Gaussian modulated data payloads and remains robust to multipath
propagation. Actually, tackling coherent sources in multipath was
among the main motivations for introducing higher-order array signal
processing \cite{yuen1997DOA}. 
\begin{figure}[t]
\centering{}\subfloat[]{\begin{centering}
\includegraphics[width=0.23\textwidth]{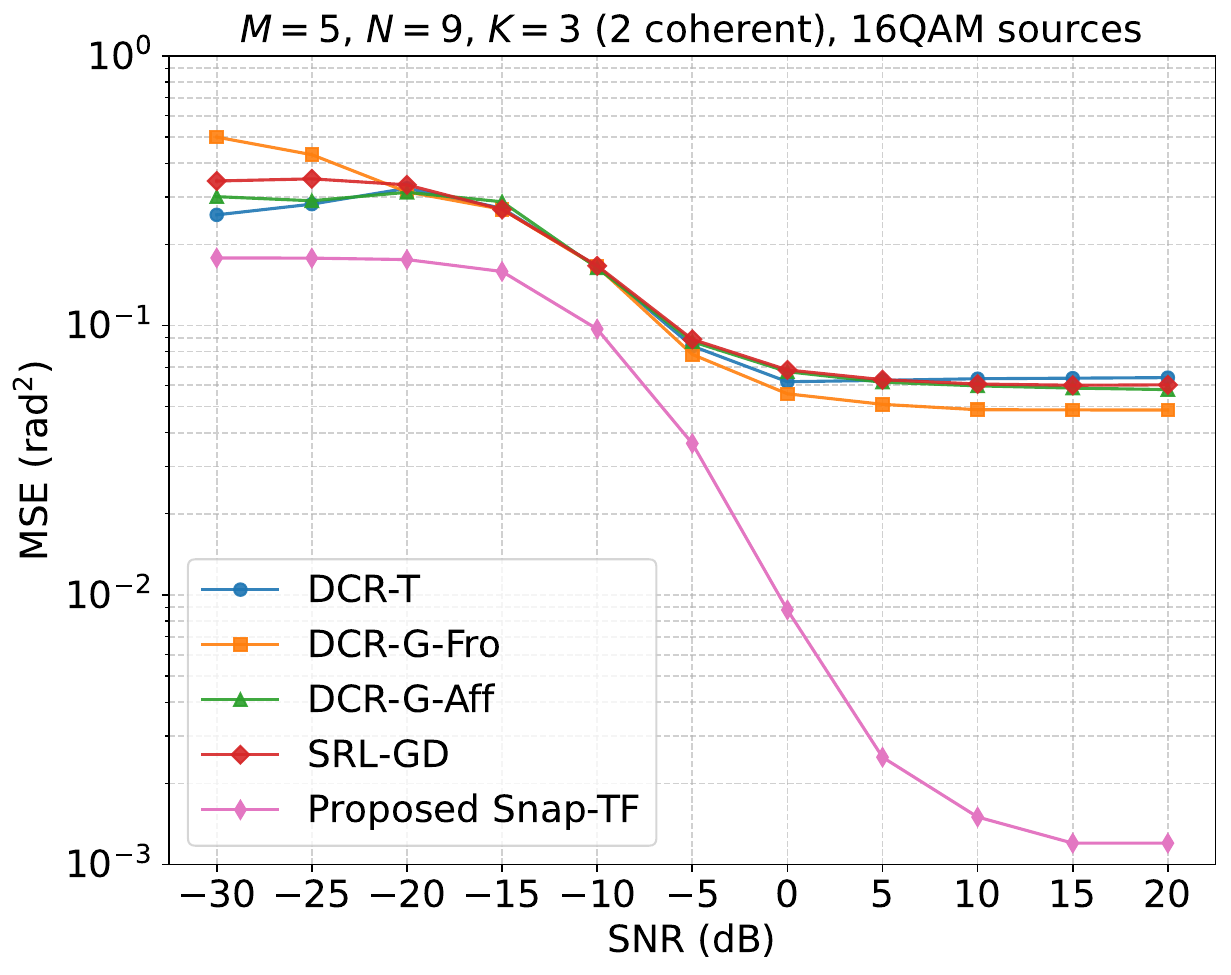}
\par\end{centering}
}\,\subfloat[]{\centering{}\includegraphics[width=0.23\textwidth]{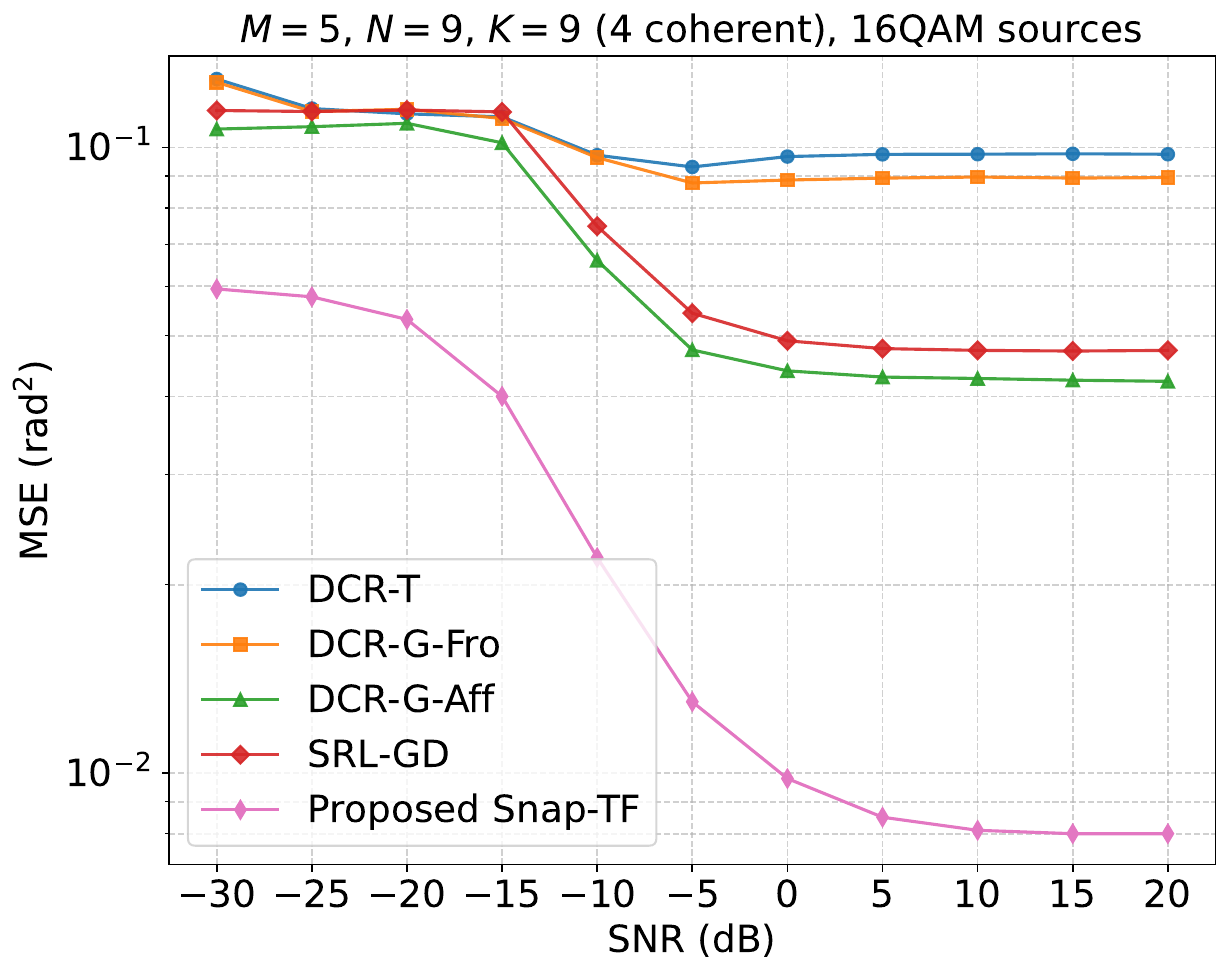}}\caption{MSE as a function of SNR under coherent sources with $M=5$, $N=9$
and 16QAM symbols.\label{fig:coherent-sources}}
\end{figure}

We present the simulation results in Fig. \ref{fig:coherent-sources}.
Specifically, Fig. \ref{fig:coherent-sources}(a) shows the case of
$K=3$ sources, where 2 of them are coherent, while Fig. \ref{fig:coherent-sources}(b)
simulates the case of $K=9$ sources, where 4 of them are coherent.
All SCM-based baselines exhibit an early performance saturation as
SNR increases over -5 dB as a consequence of coherent sources. In
addition, their performance notably degrades as compared to that under
independent sources shown in Fig. \ref{fig:MSE-vs-SNR}. By contrast,
the proposed Snap-TF algorithm shows an increasingly wide performance
advantage at high SNRs, and exhibits a similar performance to that
under independent sources, verifying its robustness to multipath propagation.

\section{Further Application: Sensing-Assisted MU-MIMO Beam Management\label{sec:Further-Application:-Sensing-Ass}}

Beam sweeping is an important step of beam management to establish
initial alignment \cite{yu2023adaptive}. In the downlink, the BS
will transmit synchronization signal blocks sequentially across various
spatial directions, effectively ``sweeping'' its coverage range.
The users then feedback the index of the strongest beams they receive
to the BS to facilitate beam alignment. In the beam sweeping stage,
the BS needs to sweep its whole coverage range based on a pre-defined
codebook, for which a widely used choice is the discrete Fourier transform
(DFT) codebook. Advanced codebook design has also been extensively
studied to reduce the beam sweeping overhead and improve the performance.

However, it is less explored how to prune the beam sweeping range
from the whole coverage area to a much smaller region during regular
data transmission. This is what we hope to achieve through the sensing-assisted
beam training. Thanks to the angular reciprocity, the uplink DOAs
and the downlink DOAs are very close even in frequency division duplex
(FDD) MU-MIMO systems \cite{xie2018channel,zhang2018directional}.
In the last section, we have shown that the proposed Snap-TF algorithm
can accurately localize the DOAs of multiple sources based on the
unknown uplink data symbols. These DOAs can help significantly narrow
down the beam sweeping candidates without additional overhead. Importantly,
while we will employ naive sweeping in the following experiments for
illustrative purposes, sensing-assisted beam sweeping is fully compatible
with any advanced beam sweeping codebook or algorithm in the literature.
The sensed DOAs help greatly reduce the search space and provides
a better initialization, making it a complementary enhancement for
existing techniques.

\subsection{Procedures}

To show the capability of our sensing-assisted beam training to both
reduce the beam sweeping overhead and improve the system throughput,
we perform simulations in MU-MIMO systems. The number of snapshots
for DOA estimation is set as $T=100$. We consider two cases where
the SNR is set as -5 dB and 5 dB in the DOA estimation stage using
unknown uplink data symbols. In the downlink transmission stage, we
set the SNR as 10 dB. The $Q$ sweeping beams are chosen from a codebook
$\mathcal{W}=\{\mathbf{w}_{q}=\mathbf{a}_{\bm{\Omega}}(\vartheta_{q})\}_{q=1}^{Q}$,
in which $\vartheta_{q}=\sin^{-1}(\frac{2q}{Q}-1)$ with $q\in\{1,2,\ldots,Q\}$,
according to Type I codebook \cite{dahlman20205g}. We set the number
of beam candidates as $Q=M$. Upon receiving the sweeping beams, each
user will feedback its strongest beam index to the BS. The coherence
interval, beam training overhead, and the feedback overhead are denoted
by $T_{\text{c}}$, $T_{\text{train}}$, and $T_{\text{fb}}$ symbols,
respectively.We evaluate the performance under $M=N=64$ with index
set $\bm{\Omega}=\{1,2,3,\ldots,64\}$ and consider $K=10$ sources.
The other parameters are the same as Section \ref{subsec:Settings}.
We compare the following methods to show the advantages of sensing-assisted
beam management.
\begin{itemize}
\item \textbf{Codebook only (full sweeping}): The BS transmits over each
of the $M$ codebook beams in sequence. All $K$ users measure these
$M$ pilot transmissions, record the strongest beam index, and feed
back to the BS. The beam sweeping overhead is $T_{\text{train}}=M$
symbols, and the feedback overhead is $T_{\text{fb}}=K$ symbols.
\item \textbf{Codebook + sensing (pruned sweeping}): For the $k$-th user,
only those candidate beams in the codebook $\mathbf{w}_{m}$ whose
steering angles satisfy $|\vartheta_{m}-\hat{\theta}_{k}|<\delta,\forall m$
are sounded, where $\hat{\theta}_{k}$ denotes the estimated uplink
DOA of the $k$-th user. Here, $\delta$ is the width of the search
window that can be freely configured. We set $\delta$ as twice of
the root MSE of the uplink DOA estimation. In this process, many beam
candidates can be pruned away. Denote $\mathcal{W}_{k}\subseteq\mathcal{W}$
as the set of beams for the $k$-th user after pruning. The beam sweeping
overhead is $T_{\text{train}}=|\bigcup_{k=1}^{K}\mathcal{W}_{k}|\leq M$
symbols, where $\bigcup$ denotes the union of sets, and the feedback
overhead is $T_{\text{fb}}=K$ symbols.
\item \textbf{Sensing only (no sweeping}): The BS directly reuses the estimated
uplink DOA $\hat{\theta}_{k}$ as the beam direction for the $k$-th
user without beam sweeping and feedback. The overheads are $T_{\text{train}}=T_{\text{fb}}=0$
symbols. 
\begin{figure}[t]
\centering{}\includegraphics[width=0.48\textwidth]{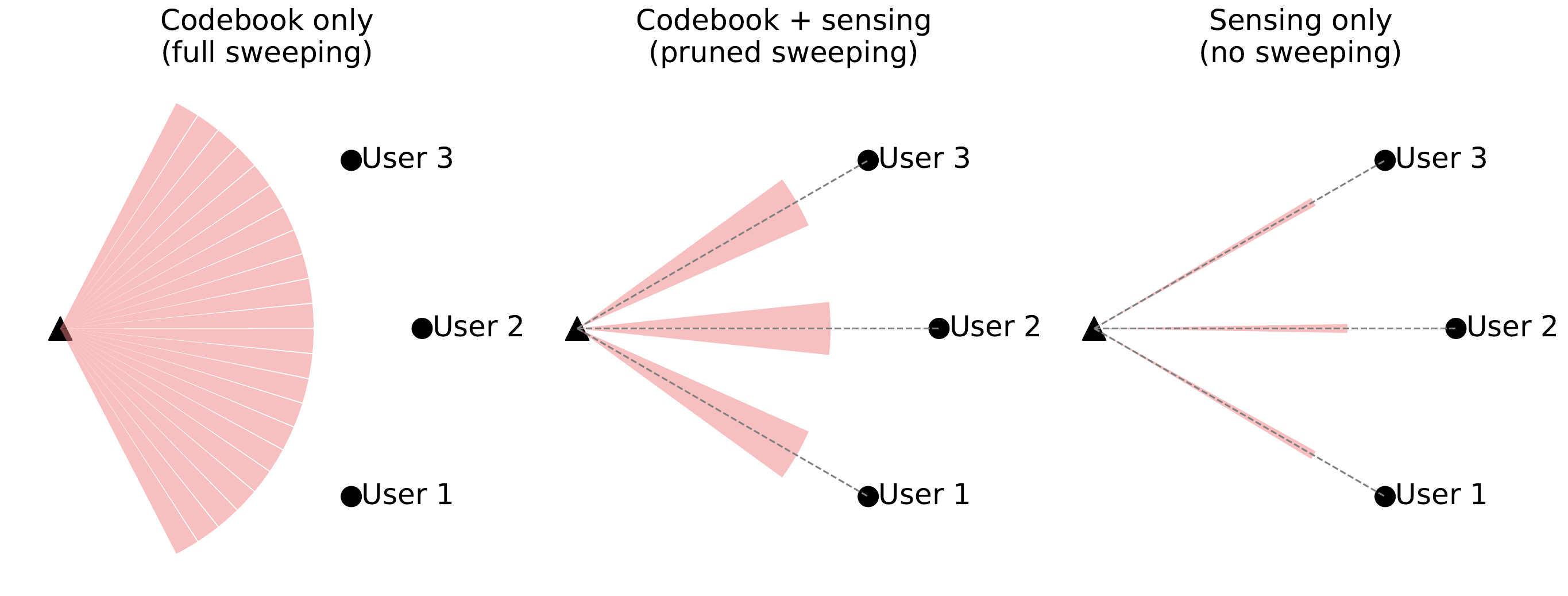}\caption{Schematic diagram of three beam training schemes, i.e., codebook only
(full sweeping), codebook + sensing (pruned sweeping), and sensing
only (no sweeping). The dashed lines in the figure denote the estimated
uplink DOAs, while the black triangles denote the BSs. \label{fig:beam-sweeping}}
\end{figure}
\end{itemize}
In Fig. \ref{fig:beam-sweeping}, we draw diagrams of these schemes.
After beam sweeping, the BS sends a small number pilots along the
selected beam directions to estimate the channel gains by the least
squares (LS) algorithm. Then, zero-forcing (ZF) precoding is adopted
for downlink transmission. Similar to \cite{zhang2018directional},
we compare the system throughput, denoted by $R=\Bigl(1-\frac{T_{\mathrm{train}}+T_{\mathrm{fb}}}{T_{c}}\Bigr)\sum_{k=1}^{K}\log_{2}\bigl(1+\mathrm{SINR}_{k}\bigr)$,
where $\text{SINR}_{k}$ is the signal-to-interference-plus-noise
ratio of the $k$-th user.

\subsection{Performance}

\begin{figure}[t]
\centering{}\subfloat[]{\centering{}\includegraphics[width=0.23\textwidth]{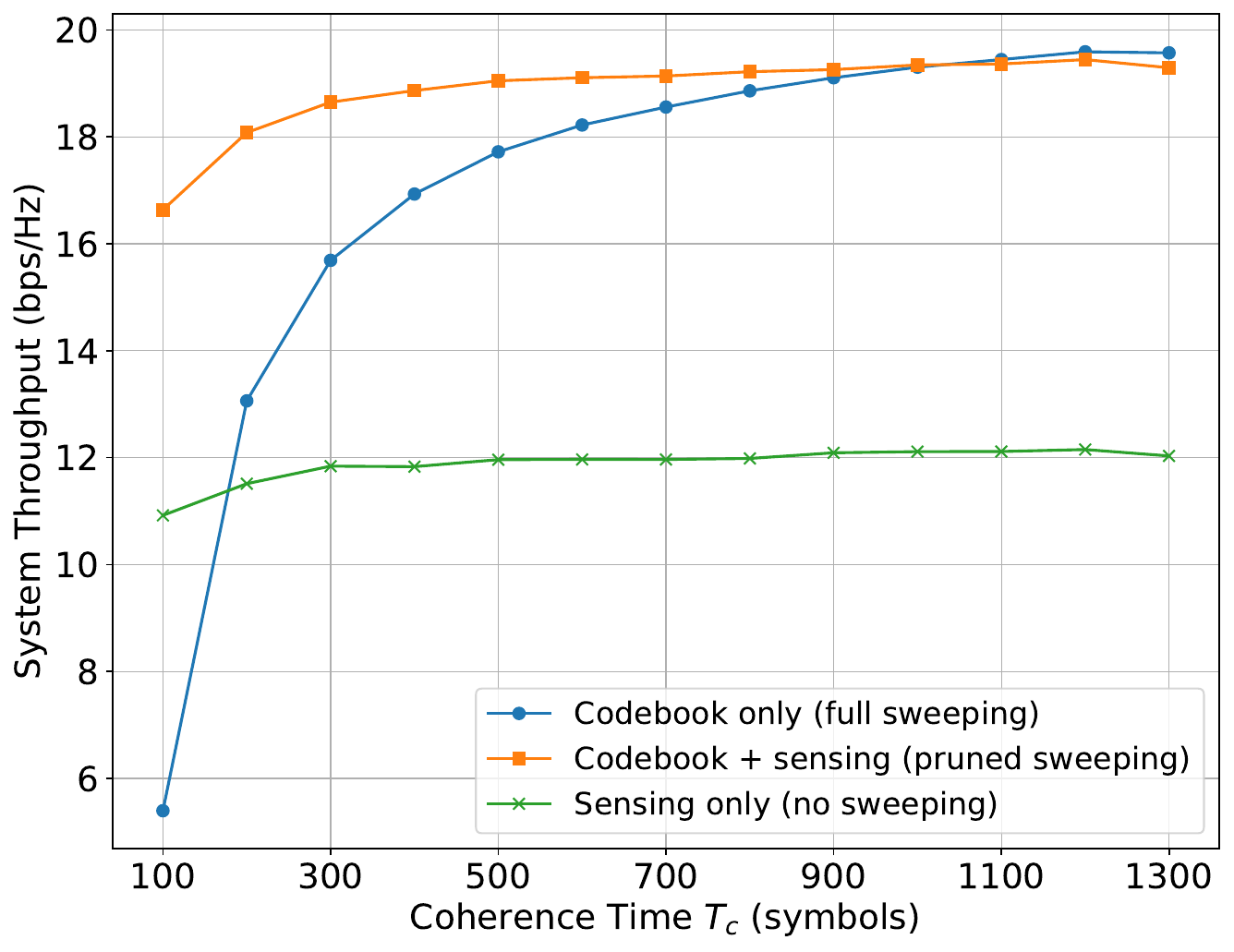}}\,\subfloat[]{\centering{}\includegraphics[width=0.23\textwidth]{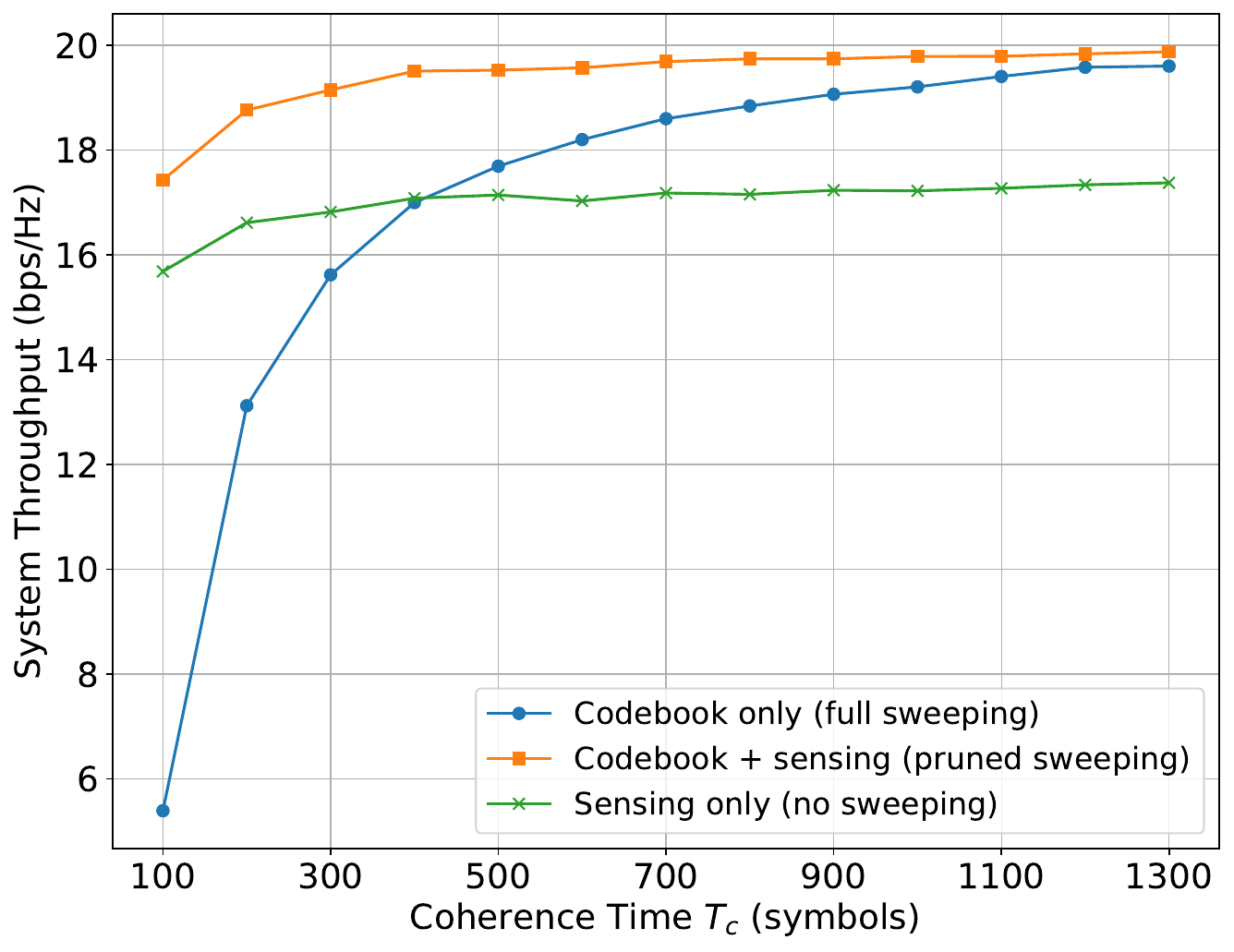}}\caption{System throughput as a function of the coherence time $T_{\text{c}}$
for different beam sweeping schemes. \label{fig:System-throughput-as}}
\end{figure}

Fig. \ref{fig:System-throughput-as} illustrates the system throughput
as a function of the channel coherence time $T_{\text{c}}$ for three
beam sweeping strategies. In Fig. \ref{fig:System-throughput-as}
(a) and (b), DOA estimation from unknown uplink data symbols is performed
at SNRs of -5 dB and 5 dB, respectively. Each data point in the plot
is averaged over 10,000 Monte Carlo trials. The proposed codebook
+ sensing (pruned sweeping) approach consistently outperforms the
conventional codebook-only method across all coherence time lengths,
with the most pronounced advantage in the short coherence regime.
This gain is attributed to the significant reduction of beam sweeping
candidates enabled by the sensed uplink DOAs, which effectively lowers
training overhead. The beam sweeping overhead for full sweeping is
$T_{\text{train}}=64$, while the overhead of sensing pruned sweeping
is significantly smaller, only leading to an average sweeping overhead
of $T_{\text{train}}=5.0$ in (a) and $T_{\text{train}}=2.9$ in (b),
which are respectively $12.8\times$ and $22.1\times$ times fewer
compared to full sweeping. As the coherence interval $T_{\text{c}}$
increases, the performance of the full sweeping baseline gradually
converges toward that of the proposed sensing-aided pruned sweeping
scheme.

Comparing the sensing-only (no sweeping) method in both subfigures,
we observe that in the low-SNR case, i.e., Fig. \ref{fig:System-throughput-as}(a),
the larger DOA estimation error renders direct reuse of the uplink
DOAs for downlink transmission infeasible, whereas in the mid-SNR
case, i.e., Fig. \ref{fig:System-throughput-as}(b), the reduced estimation
error yields better performance for downlink transmission. Although
the sensing-only scheme eliminates beam sweeping overhead, its performance
is highly sensitive to the DOA estimation accuracy. By contrast, the
codebook + sensing (pruned sweeping) scheme is much more robust, delivering
consistently competitive performance in both the low-SNR (a) and mid-SNR
(b) cases. This demonstrates that even under low SNR, the DOA estimates
are accurate enough to prune away a large number of beam candidates
for downlink beam sweeping and achieve much higher system throughput.
This demonstrates the value of the ``sensing for free'' paradigm
and the effectiveness of the proposed algorithms for improving the
communications performance. 

Importantly, the proposed sensing-assisted beam sweeping approach
works seamlessly with existing 3GPP 5G NR standards and can be easily
integrated into future 6G systems without requiring additional signaling
overhead or changes to frame structures and waveforms. The pilot-free
sensing leverages existing uplink data symbols that are part of the
standard transmission protocol, making deployment straightforward
within current and future cellular networks.

\subsection{Robustness to Reciprocity Mismatch}

In the preceding results we assumed angular reciprocity. In practice,
depending on the carrier frequency and propagation environment, slight
reciprocity mismatches between uplink and downlink DOAs may occur.
In such cases, the proposed sensing-assisted beam management method,
specifically the codebook + sensing (pruned sweeping) scheme, remains
applicable by appropriately enlarging the width $\delta$ of the search
window so that the reciprocity mismatch can be absorbed, thereby preserving
the robustness.

To validate this, we conduct a sensitivity test in which the downlink
DOA $\theta_{\mathrm{DL},k}$ is modeled the sum of the uplink DOA
$\theta_{\mathrm{UL},k}$ and a reciprocity mismatch term $\mathrm{mis}$,
given by $\ensuremath{\theta_{\mathrm{DL},k}=\theta_{\mathrm{UL},k}+\mathrm{mis}}$,
where $\ensuremath{\mathrm{mis}\sim\mathcal{N}(0,\sigma_{\mathrm{mis}}^{2})}$
is modeled as a Gaussian random variable with variance $\sigma_{\mathrm{mis}}^{2}$.
We adopt the same simulation setting as Fig. \ref{fig:System-throughput-as}(b),
except that we now fix the coherence interval as $T_{\text{c}}=200$
symbols, and sweep the strength of reciprocity mismatch between $\ensuremath{\sigma_{\mathrm{mis}}\in[0^{\circ},8^{\circ}]}$.
We compare two different search window widths $\delta\in\{5^{\circ},10^{\circ}\}$.
\begin{figure}[t]
\centering
\includegraphics[width=0.23\textwidth]{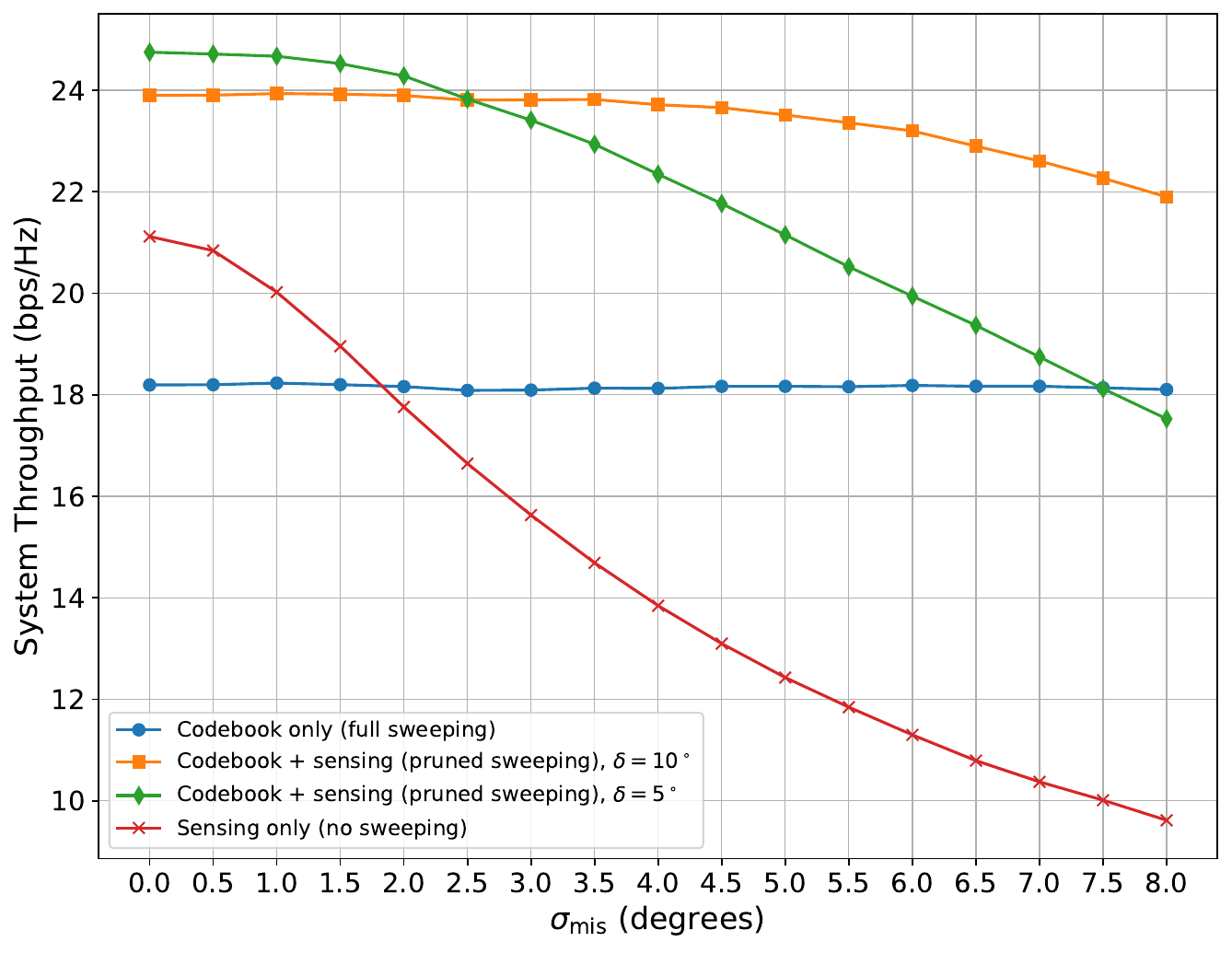}\caption{Robustness of the proposed sensing-assisted beam management to the
reciprocity mismatch of uplink and downlink DOAs. \label{fig:Robustness-of-the}}
\end{figure}

In Fig. \ref{fig:Robustness-of-the}, we plot the system throughput
as a function of $\sigma_{\mathrm{mis}}$. The results show that a
larger window, $\delta=10^{\circ}$, makes the pruned sweeping scheme
highly robust to reciprocity mismatch. The system throughput remains
essentially unchanged for $\sigma_{\mathrm{mis}}\approx0^{\circ}\thicksim5^{\circ}$,
with only mild degradation beyond this range. In contrast, narrowing
the window to $\delta=5^{\circ}$ reduces robustness. The performance
begins to decline once $\sigma_{\mathrm{mis}}$ exceeds approximately
$2^{\circ}\thicksim3^{\circ}$. These results suggest that by appropriately
setting the search window width $\delta$, sensing-assisted beam management
can robustly handle the reciprocity mismatch and outperform existing
full sweeping schemes.

\section{Conclusion and Future Work\label{sec:Conclusion-and-Future}}

This paper demonstrates the novel paradigm of ``sensing for free''—enabling
multi-source localization at the BS based on unknown uplink data symbols,
without dedicated pilots or additional overhead. Sparse arrays prove
particularly promising for this paradigm, as they have the potential
to localize more sources than antennas by exploiting their enlarged
virtual array. Our attention-only transformer directly processes raw
signal snapshots to exploit higher-order statistics for grid-less
end-to-end multi-source DOA estimation, circumventing the computational
bottlenecks and performance degradation inherent in existing DL-based
subspace methods. These accurate DOA estimates can enable significant
pruning of downlink beam sweeping candidates, and hence substantially
reduce beam training overhead and enhance system throughput at no
extra cost. By leaving frame structures and waveforms untouched, the
paradigm integrates seamlessly into existing wireless systems and
provides inspiration for next-generation 3GPP standardization efforts.
Overall, this work reveals how uplink data symbols can be harnessed
for free to unlock the substantial sensing capabilities, motivating
future research into the untapped potential of regular data symbols
for sensing.

Looking ahead, responsible deployment of ``sensing for free'' requires
addressing two important issues: resilience to potential attacks (e.g.,
location spoofing or intentional interference) and protection of the
sensitive positional information. Developing rigorous threat models,
validating standard-compliant defense strategies, and establishing
privacy-preserving practices and policies in realistic deployments
are important research directions. In addition, generalizing the proposed
algorithm to wideband and near-field systems is an ongoing future
work.

\section*{Acknowledgment}

The authors gratefully acknowledge the helpful discussions with Prof.
Xiangxiang Xu from the University of Rochester, Prof. Shenghui Song
and Mr. Ruoxiao Cao from HKUST, Mr. Xinjie Yuan from Tsinghua University,
and Prof. Lei Xie from Southeast University. We also thank Dr. Kuan-Lin
Chen from UCSD for sharing the codes of \cite{chen2025subspace}.

\bibliographystyle{IEEEtran}
\bibliography{references_JSAC}

\begin{IEEEbiography}[{\includegraphics[width=1in,height=1.5in]{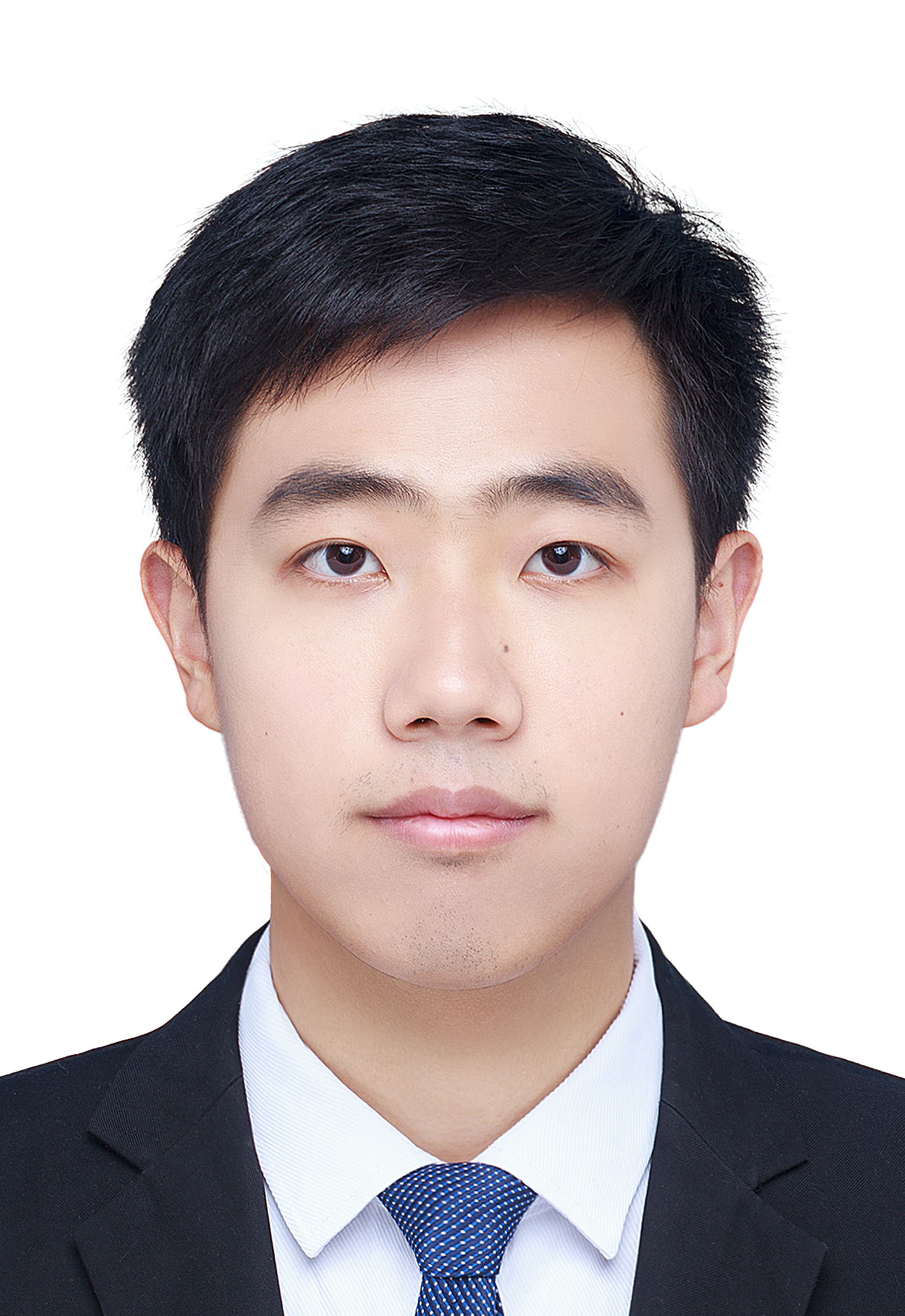}}]{Wentao Yu}
 (Member, IEEE) received the B.Eng. degree in electronic science
and engineering from Nanjing University, Nanjing, China, in 2021,
and the Ph.D. degree in electronic and computer engineering from the
Hong Kong University of Science and Technology (HKUST), Hong Kong,
in 2025. He is currently a Postdoctoral Research Fellow with the department
of electrical and computer engineering, the University of British
Columbia (UBC), Vancouver, BC, Canada. From Sept. 2024 to Jul. 2025,
he was a visiting researcher with the Research Laboratory of Electronics,
Massachusetts Institute of Technology (MIT), Cambridge, MA, USA. His
research interests lie in signal processing and machine learning for
wireless communications, including next-generation MIMO technologies,
integrated sensing and communication, and edge AI. Dr. Yu was a recipient
of the China National Scholarship in 2018 and the Hong Kong Ph.D.
Fellowship Scheme (HKPFS) in 2021. He was also recognized as an Exemplary
Reviewer of the \textsc{IEEE Wireless Communications Letters} in
2024. 
\end{IEEEbiography}

\begin{IEEEbiography}[{\includegraphics[width=1.05in,height=1.5in]{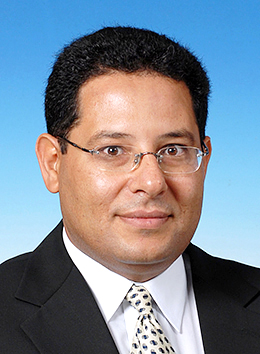}}]{Khaled B. Letaief}
 (Fellow, IEEE) is a globally recognized leader in wireless communications
and networks, with a research focus that spans artificial intelligence,
integrated sensing and communication, mobile cloud and edge computing,
federated learning, and 6G systems. His prolific contributions include
over 700 publications, which have garnered more than 64,000 citations
with an h-index of 112. He holds 15 inventions, including 11 U.S.
patents. Dr. Letaief is a distinguished member of several esteemed
organizations, including the United States National Academy of Engineering,
IEEE Fellow, and Fellow of the Hong Kong Institution of Engineers.
He is also a member of the Hong Kong Academy of Engineering Sciences.
His research excellence has earned him recognition as an ISI Highly
Cited Researcher, and he was named one of the top 30 Most Influential
Scholars in AI and the Internet of Things in 2020.

His accolades include numerous prestigious awards, such as the 2024
IEEE James Evans Avant Garde Award, 2024 Distinguished Purdue University
Alumni Award, 2022 IEEE Edwin Howard Armstrong Achievement Award,
and 2021 IEEE Communications Society Best Survey Paper Award. He has
also received the 2019 Joint Paper Award from the IEEE Communications
Society and Information Theory Society, the 2016 IEEE Marconi Prize
Award in Wireless Communications, and over 20 IEEE Best Paper Awards.

Since 1993, Dr. Letaief has been a faculty member at The Hong Kong
University of Science and Technology (HKUST), where he has held multiple
leadership roles, including Senior Advisor to the President, Acting
Provost, Head of the Electronic and Computer Engineering Department,
and Director of the Hong Kong Telecom Institute of Information Technology.
He served as Chair Professor and Dean of Engineering at HKUST and,
from 2015 to 2018, was Provost at Hamad Bin Khalifa University in
Qatar, where he played a key role in establishing a research-intensive
university in collaboration with renowned institutions like Northwestern
University, Carnegie Mellon University, Cornell, and Texas A\&M.

Dr. Letaief is celebrated for his dedicated service to professional
societies and IEEE, having held numerous leadership positions, including
Division Director and member of the IEEE Board of Directors, founding
Editor-in-Chief of the esteemed \textsc{IEEE Transactions on Wireless Communications},
and President of the IEEE Communications Society from 2018 to 2019,
the leading global organization for communications professionals.

He earned his B.S. degree with distinction in Electrical Engineering
from Purdue University in December 1984, followed by an M.S. and Ph.D.
in Electrical Engineering from the same institution in August 1986
and May 1990, respectively. In 2022, he received an honorary Ph.D.
from the University of Johannesburg, South Africa.
\end{IEEEbiography}

\begin{IEEEbiography}[{\includegraphics[width=1in,height=1.5in]{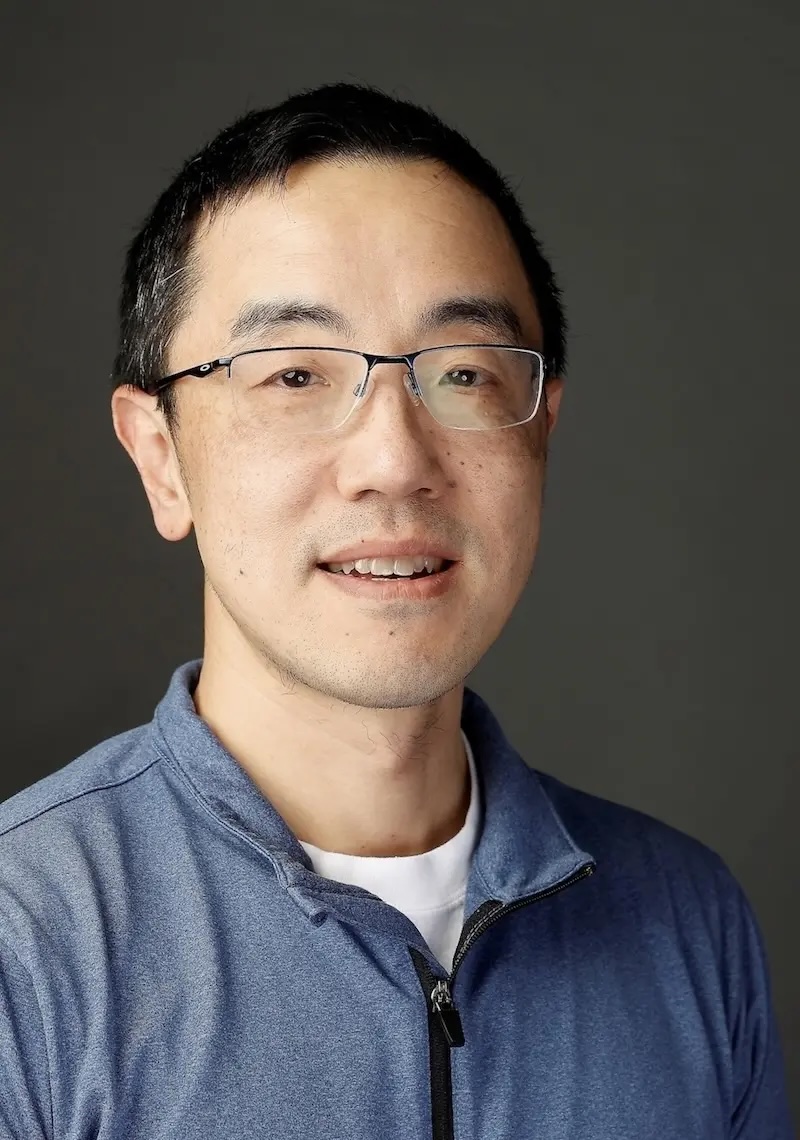}}]{Lizhong Zheng}
 (Fellow, IEEE) received the B.S. and M.S. degrees from Tsinghua
University, Beijing, China, in 1994 and 1997, respectively, and the
Ph.D. degree from the University of California at Berkeley, Berkeley,
CA, USA, in 2002. Since 2002, he has been with the Department of Electrical
Engineering and Computer Sciences, Massachusetts Institute of Technology,
Cambridge, MA, USA, where he is currently the Elihu Thomson Professor
of electrical engineering and computer sciences. He works in the general
area of information theory, statistical inference, data processing,
wireless communications, and networks. His current research interest
includes statistics, information theory, and their applications in
data science. He was a recipient of the Eli Jury Award from UC Berkeley
in 2002, the IEEE Information Theory Society Paper Award in 2003,
the NSF CAREER Award in 2004, and the AFOSR Young Investigator Award
in 2007. He is currently the Editor-in-Chief for the \textsc{IEEE Transactions on Information Theory}.
\end{IEEEbiography}

\end{document}